\documentstyle[preprint,aps]{revtex}

\newcommand{\be}{\begin{equation}}
\newcommand{\ee}{\end{equation}}

\newcommand{\nh}{$\nu h_{11/2}$}
\newcommand{\na}{$\nu i_{13/2}$}
\newcommand{\nb}{$\nu g_{7/2}$}
\newcommand{\ph}{$\pi h_{11/2}$}
\newcommand{\pg}{$\pi g_{9/2}$}
\newcommand{\hb}{(\nh)$^2$}
\newcommand{\hg}{(\nh)$^2$(\pg)$^2$}
\newcommand{\hhg}{(\nh)$^4$(\pg)$^2$}

\newcommand{\hhig}{(\nh)$^4$(\na)$^1$(\pg)$^2$}

\newcommand{\hhigg}{(\nh)$^4$(\na)$^1$(\pg)$^4$}

\newcommand{\nzpz}{$\nu 6^0\pi 5^0$}

\newcommand{\nopz}{$\nu 6^1\pi 5^0$}

\newcommand{\nopt}{$\nu 6^1\pi 5^2$}

\newcommand{\ntpo}{$\nu 6^2\pi 5^1$}
\newcommand{\ntpt}{$\nu 6^2\pi 5^2$}
\newcommand{\nrpt}{$\nu 6^3\pi 5^2$}

\tighten

\begin{document}
\draft

\title{Equilibrium shapes and high-spin properties of
the  neutron-rich $A$$\approx$100 nuclei}

\author{J. Skalski $^{1,2}$, S. Mizutori$^{1,3}$,
 and W. Nazarewicz  $^{3-5}$}

\address{
$^1$~Joint Institute for Heavy-Ion Research and
Physics Division, \\Oak Ridge National Laboratory, Oak Ridge, 
Tennessee 37831, U.S.A. \\[1mm]
$^{2}$~So\l tan Institute for Nuclear Studies,\\
ul. Ho\.za 69, PL-00681, Warsaw, Poland \\[1mm]
$^3$Physics Division, Oak Ridge National Laboratory             \\
   P.O. Box 2008, Oak Ridge,   Tennessee 37831, U.S.A. \\[1mm]
$^4$~Department of Physics and Astronomy, University of Tennessee,\\
Knoxville, Tennessee 37996, U.S.A. \\[1mm]
$^5$~Institute of Theoretical Physics, Warsaw University,\\
ul. Ho\.za 69, PL-00681 Warsaw, Poland}

\maketitle

\begin{abstract}
Shapes and high-spin properties of nuclei from the
neutron-rich ($N$$>$56) zirconium region
are calculated using the Nilsson-Strutinsky method
with the cranked
Woods-Saxon average potential and monopole pairing
residual interaction.
The shape coexistence effects and the competition
between rotationally-aligned {$1h_{11/2}$} neutron and {$1g_{9/2}$} proton
bands is discussed. Predictions are made for
the low-lying superdeformed
bands in this mass region, characterized by the intruder states
originating 
from the ${\cal N}$=5 and 6 oscillator shells.
\end{abstract}


\section{Introduction}

Nuclei from the heavy-Zr region ($Z$$\sim$40, $N$$>$56) exhibit
a wealth of structure phenomena such as, shape coexistence
effects, dramatic variations in quadrupole collectivity, strong
octupole correlations, and the existence of low-lying intruder
states \cite{[Zir88],[Woo92],[Lhe94]}.  Most nuclei from this
region are on the neutron-rich side and, therefore,  they are
not accessible using the standard in-beam techniques.  The
heaviest known isotopes in the A$\sim$100 mass region, namely
$^{105}$Y, $^{107}$Zr, $^{110}$Nb, $^{112}$Mo, $^{115}$Tc, and
$^{117}$Ru, have recently been observed \cite{[Ber94]} using the
technique of in-flight isotopic   separation of  projectile
fission fragments at relativistic velocities. Properties of
$\beta$-decay of $^{103}$Y, $^{105}$Zr, $^{110}$Nb, $^{110}$Mo,
$^{113}$Tc, and $^{115}$Ru, have been  studied
\cite{[Ays92],[Meh96]} at the IGISOL facility. These nuclei
determine the present experimental spectroscopic boundary of
this region.

A powerful tool to carry out spectroscopic studies in the neutron-rich systems
is to analyze the prompt gamma rays from nascent fission fragments.
In such measurements it is possible to
approach {\em spectroscopically} 
very neutron-rich nuclei such as $^{102}$Sr, $^{104}$Zr, $^{108}$Mo
\cite{[Hot90],[Hot91]}, and $^{112}$Ru \cite{[Lu95]}. 
Rotational structures in this region
have also been studied using breakup fusion reactions \cite{[Hae86]},
 ($t,p$) reactions \cite{[Est89]}, heavy-ion-induced
fusion \cite{[Gel88]},
incomplete fusion \cite{[Dej95]}, neutron-induced fission
\cite{[Sha94a],[Sch95a]},
and deep inelastic reactions \cite{[Reg96],[Men96]}.

In spite of many experimental efforts,
detailed information on high-spin properties of  $A$$\sim$100
nuclei is still scarce. 
The first
 systematic study of high-spin states in even-even nuclei,
 as well
as odd-$A$ nuclei in this region,
 was reported  in Ref.~\cite{[Hot91]}.
In the even-even Zr and Mo isotopes,
the yrast lines are known up to $I^\pi$=12$^+$\cite{[Dur95],[Dur96],[Smi96]}. 
In several nuclei such as $^{102}$Ru\cite{[Hae86]},
$^{104}$Ru \cite{[Reg96]}, $^{108-112}$Ru \cite{[Lu95]}, and
$^{104}$Mo \cite{[Reg96]},
a  crossing between the ground band (g-band)
 and the aligned band (s-band)
 has been  observed. Also,
in a number of  Ru  and Mo isotopes, side-bands have been 
studied 
\cite{[Lu95],[Sha94a],[Ays90],[Gue95],[Gue96]}. 
High-spin states in odd-$A$ nuclei
have been found in
$^{101}$Sr \cite{[Lhe95b]},
$^{101}$Zr \cite{[Lhe95]},
$^{103}$Zr \cite{[Lhe96]},
$^{101}$Nb \cite{[Ohm91]},
$^{103}$Nb \cite{[Lia93]},
$^{103}$Mo \cite{[Lia93a]},
$^{105}$Mo \cite{[Lia95]},
$^{103}$Rh \cite{[Dej88]},
and
$^{109,111}$Ru \cite{[But95]}.
For the heavy 
Sr isotopes, spectroscopic studies \cite{[Pfe95a],[Lhe95a]} are still 
in a preliminary stage, and practically nothing is known on the 
neighboring Kr and Se nuclei.
It is conceivable that with the help of the new generation
gamma-ray  arrays such as GammaSphere
and EuroBall, and with the help of  radioactive beam facilities,
the borders of the heavy-Zr regions will soon be expanded
towards even heavier systems.

The strong dependence of observed
spectroscopic properties on
the number of  protons and neutrons,
makes the neutron-rich $A$$\approx$100 nuclei  a very
good region  for testing various theoretical models.
According to the mean-field-based 
calculations, 
strong shape variations in this region may be
attributed to shell effects associated with large spherical
and deformed subshell closures in the
single-particle spectrum. The single-particle diagram, representative
of the nuclei from the heavy-Zr region, is shown in
Fig.~\ref{WSLevel}.
The strongest shell effects are expected for a 
spherical shape with
 the magic numbers 28, 50, and 82, and also
for the  subshell closures at $Z$=40 and $N$=56.
The transition to a deformed region occurs near $N$=58 and 76;
the strongest shell polarization towards deformed shapes is 
predicted for
$N$=60-72. Finally,
the best candidates for superdeformed shapes are nuclei
with particle numbers near 42 and 64.
[Shell-correction landscapes for  the
Zr-region are shown,
e.g., in Ref.~\cite{[Ben84]} 
(folded-Yukawa model),
Ref.~\cite{[Naz88]} (Woods-Saxon model), and 
Refs.~\cite{[Rag84],[Rag88]} (Nilsson model).]

According to calculations based on the mean-field approach,
the occupation of {$1h_{11/2}$} neutron  and {$1g_{9/2}$} proton orbitals   is essential for
 understanding the  deformed configurations around $^{100}$Zr \cite{[Naz88a]}. 
The best examples of shape coexistence in this region are Sr, Zr, and
Mo isotopes with $N$$\sim$58.
In the language of the deformed shell model, the onset
of deformation around $N$=58 can be associated with the 
competition between the spherical gaps at $Z$=38, 40, and  $N$=56, 
and the deformed subshell closures at
$\beta_2$$\approx$0.35 at particle numbers $Z$=38, 40, and 
$N$=60, 62, and 64.
Theoretically, the delicate
energy  balance between spherical and deformed configurations depends
crucially on the size of these gaps in the single-particle 
energies.
As discussed in Refs.~\cite{[Dob88],[Dob88a],[Wer94]}
the deformation onset at
$N$$\approx$58 results from  the subtle
interplay between the deformation-driving neutron-proton energy (varying
smoothly with the shell filling) and the symmetry-restoring monopole
energy responsible for the shell effects. 

Equilibrium deformations and moments, potential energy surfaces,
microscopic structure of coexisting
configurations, and shape transitions in the heavy-Zr region 
have been calculated by many authors 
\cite{[Ben84],[Naz88],[Rag84],%
[Rag88],[Naz88a],[Dob88],[Dob88a],%
[Wer94],[Ars69],[She72],[Fae74],%
[Fed78],[Fed78a],[Azu79],[Cam80],[Mol81a],[Kho82a],[Abe82],[Hey84a],%
[Kum85a],[Bon85],[Cas85c],[Gal86],[Hey88],%
[Sha88a],[Sug90],[Que90],[Tro91],%
[Dej92],[Cha91],[Abo92],[Kir93],[She93],[Hir93],%
[Ska93],[Buc94],[Bha94],[Bha94a],[Bar95a],[Mol95],%
[Laz95],[Gia95],[Lon95],[Hey95],[Cos96],[Cos96a],%
[Sin96],[Tro96],[Dev96],[Dev96a],[Rag96]}.
In most cases,  calculations show large deformations in
Sr, Zr, and Mo isotopes with $N$$\geq$60. However, the
details of the  shape
transition near   $N$=58 is predicted differently by
various models, the onset and rapidity of this transition being very
sensitive to the  model \cite{[Woo92]}.

One of the most interesting features of nuclei from the $A$$\sim$100
region is the richness of various structural effects that occur at 
high angular momenta. Many
of these effects have a straightforward interpretation in terms of the
interplay between deformation, pairing, and the Coriolis force.
In the deformed nuclei near  $^{102}$Zr, the alignment pattern is
expected to be rather simple. 
While the first proton crossing can be associated with the breaking of
the {$1g_{9/2}$} pair, the 
lowest neutron crossing is due to the {$1h_{11/2}$} alignment.

For even-even systems,
band crossings have been seen in the g-bands of $^{108-112}$Ru 
\cite{[Lu95]}, the
$N$=60 isotones $^{104}$Ru \cite{[Hot91],[Hae86],[Reg96],[Men96]}
and $^{106}$Pd \cite{[Gra76]}, the 
$N$=58 isotones $^{100}$Mo \cite{[Hot91],[Reg96]}, 
$^{102}$Ru \cite{[Hae86]}, and  $^{104}$Pd \cite{[Gra76]},
and  the $N$=56 isotone $^{100}$Ru \cite{[Voi76]}.
For proton numbers $Z$$\ge$46 the interplay between $1g_{9/2}$ protons and 
$1h_{11/2}$ neutrons is well established \cite{[Fra85],[Kel85]}.
In the case of $^{102}$Ru, the $1h_{11/2}$ neutron crossing is
supported by blocking arguments in neighboring odd-$N$ and odd-$Z$
nuclei \cite{[Hae86],[Dej88]}.
Namely,  in the 
$N$=58 nucleus $^{103}$Rh there is a clear crossing
in both $\pi$=$-$ and  $\pi$=+ 1-q.p. proton bands, as well as in the  1-q.p. 
$\pi$=+ 
bands in $^{101,103}$Ru. 
However, no crossing has been observed  in the $\pi$=$-$,
$1h_{11/2}$  bands in $^{101,103}$Ru.  
Similarly, the presence of a band crossing                        
in  the yrast line of even-even nuclei $^{104}$Ru and 
$^{104}$Mo\cite{[Reg96]}, and its absence
in the $\pi$=$-$ 1-q.p.
bands in the neighboring odd-$N$ systems, suggest
that  {$1h_{11/2}$} neutrons are involved.

In addition to blocking arguments, there exists
other  experimental evidence
for the presence of the
{$1h_{11/2}$} neutron unique-parity states in this mass region,  notably
the {\em direct}  observation of negative-parity, 
very collective bands in $^{101,103}$Zr, assigned to the 
[532~5/2] ({$1h_{11/2}$})
Nilsson orbital \cite{[Hot90],[Dur95],[Lhe95]}.

In the Ru and Mo isotopes with $N$$\ge$60, the role of 
(static or dynamic) triaxial shape degrees of freedom
 has been discussed.
In $^{104-108}$Mo and $^{108-112}$Ru, the excitation energies
of the $2^+_\gamma$ bandheads are  low. 
Furthermore, in the $^{104,106}$Mo isotopes,
the band built upon the excited $I^\pi$=$4^+$ state,
a candidate for the double-$\gamma$ vibrational excitation, 
has been  observed \cite{[Gue95],[Gue96]}.
On the other hand, based on 
the energy differences and the $E2$ branching ratios
analysis, triaxial deformations have been suggested for
 $^{108-112}$Ru 
\cite{[Sha94a]}.
The development of triaxiality at high spins has been addressed by
measurements of
$\gamma$-band$\rightarrow$g-band branching ratios
\cite{[Lu95]} and by lifetime measurements \cite{[Dur96],[Smi96]}.

Little is known about  shapes of lighter elements in this region.
Recent laser  
measurements of isotope shifts in the heavy
Kr isotopes \cite{[Kei95],[Lie96]}
indicate the absence of a
sudden shape transition around $N$=60.

The main objective of this study is to investigate  the
importance of shape changes and pairing correlations in the
description of high-spin bands in nuclei near  $^{102}$Zr. 
Predictions have been made for those nuclei  which  show the
best prospects for triaxial shapes and for superdeformed bands 
at  relatively  low  excitation energies.

The paper is organized as follows.
The Woods-Saxon model and the total routhian surface 
technique are described in Sec.~\ref{MODEL}.
The results of ground-state calculations
in the $A$$\sim$100 region are given in Sec.~\ref{GS}. In particular,
ground-state equilibrium deformations and their dependence on
the treatment of pairing correlations and macroscopic energy are discussed.
Cranking calculations
are  presented  in
Sec.~\ref{HS}, together with the
analysis of typical alignment patterns, band structures, and shape
changes.
The summary and conclusions are contained in Sec.~\ref{CONCLUSIONS}.

\section{The model}\label{MODEL}

The
macroscopic-microscopic  method employed in this work
 is an approximation to the HF approach \cite{[Str67],[Bra72]}.
Its main assumption  is that the total energy
of a nucleus can be decomposed  into two parts,
\be\label{Eshell1}
E = E_{\rm macro}+E_{\rm micro},
\ee
where $E_{\rm macro}$
 is the macroscopic energy
and $E_{\rm micro}$ is the microscopic energy (shell correction)
calculated from a non-self-consistent average deformed potential.

The shell corrections were obtained using
the deformed Woods-Saxon (WS) potential 
of Ref.~\cite{[Cwi87]}. For the macroscopic energy,
two energy formulas were employed: the liquid-drop (LD)
model of Ref.~\cite{[Mye69]}
and the Yukawa-plus-exponential  model (finite-range liquid-drop (FRLD) model)
of Refs.~\cite{[Kra79],[Mol81]}.
The latter model is softer to shape distortions and, consequently, 
is expected to favor more deformed nuclear shapes 
\cite{[Naz84],[Nil95],[Jon96]}.

The nuclear surface was  defined by means of 
the standard multipole expansion,
\begin{equation}
R(\Omega ) = c(\alpha)R_0\left[
1 + \sum_{\lambda\mu} \alpha_{\lambda\mu}Y_{\lambda\mu}
(\Omega )\right],
 \end{equation}
with $c(\alpha)$ being determined from the volume-conservation condition and
$R_0=r_0A^{1/3}$.
[Axial deformations $\beta_\lambda$ are defined as
$\beta_\lambda\equiv\alpha_{\lambda 0}$.]

In our analysis of equilibrium shapes we did not consider the
odd multipolarity deformations.  We are mainly interested in
the dominant features related to the leading quadrupole effects.
Secondly, as discussed previously \cite{[Naz84],[Naz90b]}, the
static octupole deformations are not expected to be present in
well-deformed neutron-rich $A$$\sim$100  nuclei.

The shell correction was calculated according to
the  prescription given in Ref.\ \cite{[Bra72]}. 
The results of the calculations presented  in  this
paper  were
obtained with a  smoothing width  of 49.2/$A^{1/3}$ MeV and  a
sixth-order curvature correction.
All  single-particle states lying below a cut-off
energy of 131.2/$A^{1/3}$ MeV above the Fermi
level were included.
Our calculations are  based on a
three-dimensional minimization on the mesh, and 
the variation of the average potential with 
$Z$ and $N$ has been suppressed (i.e.,
the calculations were performed using single-particle
energies of a representative central nucleus).
 The related errors in energy surfaces introduced by this procedure 
 are  known to be small \cite{[Naz90]}.

In the  $I$=0 calculations,
pairing energies were computed using either the standard BCS treatment
or the particle number projection method before variation (PNP).
The latter method is expected to be more precise and reliable
in the regions of low single-particle level densities where the
BCS treatment breaks down.
For the pairing calculations we took
the lowest $Z$ or $N$  single-particle
orbitals  for
protons or neutrons, respectively, counting from the bottom of 
the potential well.
The pairing strengths were  determined using
the average gap method. 
The optimal choices for the 
average gap parameterization, discussed in Ref.~\cite{[Mol92b]},
are
adopted for  this study.
Namely, in the PNP variant we used  average gaps of
$\tilde\Delta_n = 8.6/\sqrt{A}$\,MeV and 
 $\tilde\Delta_p =9.9/\sqrt{A}$\,MeV, and in the BCS variant we used
 $\tilde\Delta_n =13.3/\sqrt{A}$\,MeV
 and $\tilde\Delta_p = 13.9/\sqrt{A}$\,MeV.
  
The parameters of the  WS potential were taken 
from  Ref.~\cite{[Naz88]}; however, they were adjusted 
  to describe the rapid 
 shape transition  between $^{96}$Zr and $^{100}$Zr. These parameters differ 
 from the standard set \cite{[Dud81]}
 only in the values of the spin-orbit potential strength. 
Namely, we adopted
the values   
$\lambda_{\rm so, n}$=39 and $\lambda_{\rm so, p}$=32 instead of the standard
values $\lambda_{\rm so, n}$=35 and $\lambda_{\rm so, p}$=36, respectively.

In order to study the variations of 
potential energy surfaces at $I$=0
(Sec.~\ref{GS}),
 three sets of calculations were performed:
\begin{description}
\item[V1]
Axial calculations in a three-dimensional
deformation space ($\beta_2$, $\beta_4$, $\beta_6$).
They were carried out by setting up 
a three-dimensional mesh with the mesh steps equal to  0.05 
0.04, and  0.04 for   $\beta_2$, $\beta_4$, and $\beta_6$, respectively.
In these calculations we employed both LD and FRLD models for the 
macroscopic
energy and both pairing models, i.e., BCS and PNP.
\item[V2]
The explicit energy
minimization 
in a three-dimensional
deformation space ($\beta_2$, $\beta_4$, $\beta_6$)
performed for
each isotope separately
 (here no assumption about a ``central nucleus" has been made).
In these calculations, we used the set of
WS parameters of Ref.~\cite{[Naz88]},
 the FRLD macroscopic energy, and the Lipkin-Nogami (LN)
 approximation to the PNP method (with the same pairing 
strengths as for the
PNP mesh calculations). The details of the LN treatment 
are given in Ref. \cite{[Naz90]}.
\item[V3]
Triaxial calculations in a three-dimensional
deformation grid  ($\beta_2$, $\gamma$, $\beta_4$).
 The mesh consisted of 10 points
[0.05 (0.05) 0.50] in the 
$\beta_2 \cos(\gamma+30^\circ)$ direction, 
 12 points  {[$-0.20$ (0.05) 0.35]}
in the $\beta_2 \sin(\gamma+30^\circ)$ direction,
 and 4   points in the direction of $\beta_4$. 
To  optimize the calculations,  the $\beta_4$-grid  was defined relative
to   the value of $\beta_4$
minimizing the LD energy at given 
 $\beta_2$ and $\gamma$,
$\beta_{4, {\rm LD}}(\beta_2,\gamma)$ \cite{[Naz85]}. 
Namely, the hexadecapole mesh was defined as
$\beta_{4, {\rm LD}}(\beta,\gamma) +\Delta 
 \beta_4$, $\Delta \beta_4$ being equal to --0.04 (0.04) 0.08.
The hexadecapole deformation 
 was included in a way which guarantees the modulo $60^\circ$  
 invariance of the nuclear shape in the $(\beta_2, \gamma)$ plane, 
 i.e., the nonaxial hexadecapole deformations are present for 
 $\gamma$ not equal to multiples of 60$^\circ$ \cite{[Naz84]}. 
In this variant we employed the LD model for the macroscopic
energy and both pairing models, i.e., BCS and PNP.
\end{description}

The high-spin behavior ($I$$>$0)
of $A$$\sim$100 nuclei was investigated
using the total routhian surface (TRS) approach 
of Refs. \cite{[Naz87],[Naz89]}.
 The TRS cranking calculations were performed on a three-dimensional 
deformation 
 mesh including  $\beta_2$, $\gamma$,
 and $\beta_4$ deformations. 
 The mesh was the same as that described
 in variant {\bf V3}.
The calculations were carried out  for all nuclei  with $Z$=38-44 and 
 $N$=58-66, and for
rotational frequencies $\hbar \omega$=0.075 (0.075) 1.50 MeV.
The total routhians were defined as:
   \begin{equation}
   \label{e101}
  E^{\omega}= E^{\omega=0} + 
    \langle \Psi^{\omega} \mid \hat{H}^{\omega} \mid \Psi^{\omega}\rangle -
    \langle \Psi^{\omega=0} \mid \hat{H}^{\omega=0} \mid \Psi^{\omega=0}\rangle,
 \end{equation}
 where $E^{\omega=0}$ is the energy calculated at a given deformation 
 according to the shell-correction method, 
$\hat{H}^{\omega}$=$\hat{H}$--$\omega \hat{j}_x$
 is the total routhian including the pairing term, and 
 $|\Psi^{\omega} \rangle $ is the 
independent quasi-particle 
vacuum at a given rotational frequency. 
  Due to the size   of this 
computation,
the LD model was employed  for the macroscopic energy. In our analysis
of rotational bands, we limited ourselves to the vacuum configurations of 
even-even systems. That is, only the positive-parity, even-spin
(signature $r$=1)
bands were considered.

  The average BCS pairing gap was parameterized in a way which facilitates 
 separate calculation for neutrons and for protons:  
$\tilde \Delta_n=4.93 N^{-1/3}$ 
 and $\tilde \Delta_p=4.70 Z^{-1/3}$ \cite{[Mol92b]}. 
 For nonzero spins the pairing gaps were parameterized as functions
of rotational frequency:
\begin{equation}
\Delta (\omega) =  \left \{ \begin{array}{ll}

\Delta (0) \left[1 - \frac{1}{2} \left(\frac{\omega}{\omega_c}
\right)^2\right], &
\omega < \omega_c \nonumber \\
\frac{1}{2} \Delta (0) \left(\frac{\omega_c}{\omega}\right)^2, & \omega >
\omega_c
\end{array} \right.
\end{equation}
The parameter $\hbar \omega_c$  was fixed at 0.75 MeV for neutrons
and 0.9 MeV for protons.
The proton and neutron  chemical potentials of the RBCS equations 
were  adjusted at each deformation and rotational frequency 
separately so as to satisfy the particle number equation.

\section{Ground-state equilibrium deformations}\label{GS}

The potential energy surfaces (PES)  for the neutron-rich even-even Kr-Pd 
isotopes obtained in the
axial variant {\bf V1}   
are shown in 
Figs.~\ref{PECKrZr} and \ref{PECMoPd}.
The results of triaxial
calculations (variant {\bf V3}) 
are displayed for selected nuclei in
Figs.~\ref{PESZra} ($^{100,102}$Zr), ~\ref{PESZrb} ($^{104,106}$Zr), 
~\ref{PESMoa} ($^{100,102}$Mo), ~\ref{PESMob} ($^{104,106}$Mo),
~\ref{PESRua} ($^{104,106}$Ru),
 and ~\ref{PESRub} ($^{108,110}$Ru).

 The calculated equilibrium deformations obtained by means of 
the explicit minimization at axial shape (variant {\bf V2})
are shown  in Table~\ref{table_prolate-oblate}, and  
Table \ref{table_threeDminima} displays  the calculated equilibrium 
deformation obtained 
 by a  minimization in the ($\beta_2,\gamma$)-plane
(variant {\bf V3}).

\subsection{General  properties of 
potential energy surfaces}

The modification of the  WS parameters
according to Ref.~\cite{[Naz88]} has some impact only 
on the isotopes with $N$=56 and $N$=58; the optimized
parameters tend to lower  the PES around the spherical
point. The most pronounced differences
are obtained for the Sr and Zr isotopes where the equilibrium deformations 
  are drastically changed.
 For other $N$=56, $58$ isotones the equilibrium 
 deformations are only slightly modified.
More pronounced is the dependence on the 
choice of the
macroscopic energy and on the  pairing treatment.
Namely, the most pronounced minima 
 and the largest equilibrium deformations were obtained
in the FRLD/PNP model (Figs.~\ref{PECKrZr} and ~\ref{PECMoPd}).
The LD/BCS variant 
yields rather shallow  minima, especially for the Mo and Ru isotopes, and
 the smallest equilibrium deformations.

{}For most nuclei considered, the calculations predict two 
minima corresponding to
  prolate and oblate shapes. In the Kr, Sr, and Zr  isotopes,
these minima are separated by a relatively high 
spherical  barrier. 
The largest  prolate deformations cluster
at the middle of the shell, i.e.,
around $Z$=38, 40, and $N$=60, 62, and 64. As discussed above, 
these particle numbers can be associated with  
large prolate gaps in the single-particle
spectrum (see Fig. \ref{WSLevel}).

The barrier height (and the oblate-prolate) energy
difference decreases with $Z$. As will be discussed below,
in many cases the secondary minima are unstable to triaxial deformation.
In the PNP calculations,  there appear local
substructures in the potential energy curves; the PES's
calculated in the BCS variant are  smoothed. The 
 largest effects occur in the Mo isotopes where the prolate minimum 
splits into two components, and in the Ru isotopes with
 $N$=72-76, 
where secondary minima with large deformation 
$\beta_2 \approx$0.45-0.50 appear.

In triaxial calculations ({\bf V3}),
the Sr and Zr isotopes are predicted  to be axial, while
the Mo-Pd isotopes are calculated  to be $\gamma$-soft.
This is illustrated in the potential energy surface (PES) described in
Figs~\ref{PESZra}, ~\ref{PESZrb}(Zr), ~\ref{PESMoa}, ~\ref{PESMob} (Mo), 
~\ref{PESRua} and ~\ref{PESRub} (Ru).
In the PNP variant, the PES's of the  Mo 
isotopes with $N$=58, 60 show very shallow, 
$\gamma$-unstable valleys 
corresponding to $\beta_2$ ranging from 0.22 to 0.30.
The heavier Mo isotopes with $N$=62-66 have 
 shallow triaxial minima at $\gamma$$\approx$20$^\circ$ 
and $\beta_2\approx$0.32.
All Ru isotopes have been found to be triaxial in their ground states,
with shallow minima at $\gamma$$\approx$20$^\circ$. 
Again, 
 the minima calculated  in the PNP variant are 
more pronounced
than those in the BCS variant. (For more discussion 
on the influence of the PNP on the rigidity of the PES minima,
see Refs. \cite{[Que90],[Rei96]}.)

\subsection{Equilibrium deformations and quadrupole moments}

 The calculated equilibrium deformations obtained by means of 
the explicit minimization at axial shape (variant {\bf V2})
are displayed in Table \ref{table_prolate-oblate}, 
 together with the  predicted oblate-prolate energy differences, and
  calculated and experimental~\cite{[Ram87]} intrinsic quadrupole moments.
The oblate-prolate energy difference is  defined as 
$\Delta E_{\rm op}$=$E_{\rm o}$--$E_{\rm p}$, where 
$E_{\rm o}$ corresponds to the minimum with negative
 $\beta_2$
while $E_{\rm p}$ corresponds to the minimum with zero or positive
$\beta_2$.
The quadrupole moments were
extracted from the calculated 
equilibrium deformations using the procedure outlined in Ref.~\cite{[Naz90]}.
In Table \ref{table_threeDminima}, the calculated equilibrium deformations obtained 
by means of the minimization in the ($\beta_2,\gamma$)
plane are displayed, together with the calculated 
transition quadrupole moments, defined  as \cite{[Rin82]}
\begin{equation}\label{Q2}
Q(\gamma) = \frac{2}{\sqrt{3}} Q_0\cos(\gamma + 30^\circ),
\end{equation}
where $Q_0$=$Q(\gamma=0^\circ)$.

It is instructive to compare the equilibrium deformations 
obtained in our work with those
of  Chasman calculated using the LD model and the 
shell correction approach with  WS potential
 \cite{[Cha91]}, those obtained 
by M\"oller {\it et al.} in the 
shell correction approach with the folded-Yukawa single-particle potential 
and the  finite-range droplet-model
(FRDM) \cite{[Mol95]}, and with the results of the extended-Thomas-Fermi
model (ETFSI) of Ref.~\cite{[Abo92]}.

{\em The heavy Se isotopes} are predicted to be oblate-deformed with
$\beta_2$$\approx$--0.28 and  very large values of $\beta_6$$\approx$0.10.
The prolate-deformed minima appear  0.4--1.4\,MeV higher in energy.
This result is consistent with
the ETFSI calculations which yield oblate shapes with $\beta_2$=--0.31
for all Se isotopes with $N$$>$58. On the other hand,
FRDM favors
prolate shapes with
0.29$<$$\beta_2$$<$0.33.

{\em The heavy Kr isotopes} are predicted to be oblate-deformed 
($\beta_2$$\approx$--0.31, $\beta_6$$\approx$0.10) up to $N$$\sim$66.
For the Kr isotopes with $N$$>$64, $\Delta E_{\rm op}$ is
rather small.
This is due to the large oblate  gap with $Z$=36.
This result is consistent with
the ETFSI calculations which yield an oblate-prolate
shape transition at $N$$\approx$68.
The FRDM  favors well-deformed prolate ground states with
$\beta_2$$\approx$0.33.
The recent laser-spectroscopy measurements\cite{[Kei95],[Lie96]}
show that
the observed $\langle r^2 \rangle$ does not drastically change at $N$=60,
indicating that there is no sudden increase of $|\beta_2|$ for
$N$$\le$60 nuclei.  Since the deduced $\beta_2$ is $\approx$0.3 for $N$=60,
all these calculations are  consistent with this experiment.

{\em The heavy Sr isotopes} 
with 60$\leq$$N$$\leq$72
are predicted to have
well-deformed  prolate ground states in all the models discussed.
The calculated quadrupole deformations $\beta_2$
range from 0.30 to 0.40. Contrary to the FRDM results,
both our calculations and those of the   ETFSI model
yield a rapid  shape transition
at  $N$$\approx$56.

{\em The heavy Zr isotopes} 
with 60$\leq$$N$$\leq$72
are predicted to have
well-deformed  prolate ground states 
(0.33$\leq$$\beta_2$$\leq$40)
in  all the models discussed.
At  $N$$\approx$74,
the transition to oblate shapes 
is predicted. This transition 
appears much earlier, at $N$=66, in the calculations by Chasman.
The rapid shape change 
between $N$=56 and $N$=60 is reproduced by
our model, by Chasman,  and by the ETFSI model. 
In the FRDM, the shape change is
calculated to be  more gradual; the nucleus
$^{96}$Zr is predicted to be prolate-deformed with $\beta_2$=0.2
(see the discussion in Ref.~\cite{[Ben84]} regarding the single-particle
structure of the folded-Yukawa potential near  $^{96}$Zr). 
As seen in Figs.~\ref{PESZrb},
the heavy-Zr isotopes with $N$$\ge$66  become
$\gamma$-soft.

In {\em the heavy Mo isotopes},  the calculated oblate-prolate energy
difference decreases, hence  the competition between low-lying
coexisting minima is more
pronounced. 
In contrast to the Sr and Zr isotopes,  
the transition from spherical
to deformed shapes in the Mo isotopes at $N$$\approx$58 is predicted to be
  gradual, in agreement with experimental 
findings. 
The transitional nucleus $^{102}$Mo is very soft and
the isotopes $^{104,106,108}$Mo are calculated  to have triaxial 
ground-state minima.
Similar results were predicted by Chasman, in the 
Nilsson-Strutinsky calculations of Ref.~\cite{[Gal86]}, and in  the
Skyrme-Hartree-Fock study \cite{[Bon85]}. 
The prolate-to-oblate shape transition is expected at $N$$\approx$68-70
in all these  models.

In {\em the heavy Ru isotopes}, the prolate-to-oblate shape transition is
predicted  at $N$$\approx$66 in our axial calculations and in the FRDM, but
much earlier, at $N$$\approx$60,
in the ETFSI model. Again, as in the Mo  case, 
our calculations for the Ru isotopes  predict $\gamma$-softness or
triaxiality. 
Similar results were obtained by Chasman, by the 
Nilsson-Strutinsky calculations of Ref.~\cite{[Abe82]}, and  by the
Skyrme-HF study \cite{[Ays90]}.

Stable triaxial shapes in $^{108-110}$Ru have recently been reported 
in Ref.~\cite{[Sha94a]}. 
Our calculations are consistent with  the 
experimentally deduced $\gamma$ values, namely
23$^\circ$ for $^{108}$Ru and 24$^\circ$ for $^{110}$Ru, but fail 
in reproducing  the experimental quadrupole moments. That is, 
after correcting for the effect of triaxiality using  Eq.~(\ref{Q2}),
theoretical quadrupole moments for Mo and Ru isotopes are 
predicted to be
too large.
 Interestingly, 
axial calculations reproduce measured values 
 rather well (cf. Tables \ref{table_prolate-oblate} and 
\ref{table_threeDminima}),
but they
overestimate the experimental quadrupole moments for other systems
predicted to have near-axial shapes.
A possible source of this discrepancy is the shape mixing
phenomenon 
which gives rise to the 
fragmentation of the experimental low-lying  $B(E2)$ strength.

\section{High-spin calculations}\label{HS}


\subsection{Total Routhian Surfaces: Examples}

Figures \ref{TRS100Sr}, \ref{TRS102Zr}, and \ref{TRS108Ru}
show  typical examples of total Routhian surfaces for nuclei from
the heavy-Zr region.
The nucleus $^{100}$Sr,
Fig.~\ref{TRS100Sr}, represents
a well-deformed collective rotor.
The prolate minimum, $\beta_2$$\approx$0.35,
  does not change
significantly with angular momentum up to $I$$\approx$36.
Only at a very high rotational frequency does
this minimum become triaxial ($\gamma$$>$0$^\circ$) due to alignment of
several high-$j$ quasi-particles (see Sec.~\ref{HSdef}).
The coexisting oblate minimum ($\gamma$$\approx$--60$^\circ$) 
is also fairly stable with $I$.

In $^{102}$Zr (Fig.~\ref{TRS102Zr})
 the situation is different.
There is a shape change from the 
well-deformed prolate minimum to a triaxial minimum
with $\gamma$$<$0$^\circ$  associated with  the first
band crossing.
The local minimum at $\gamma$$\approx$--60$^\circ$,
 seen in the
$\hbar\omega$=0.3 \,MeV surface, 
 quickly becomes  non-yrast with increasing $\omega$.
At very high rotational frequencies, the superdeformed 
($\beta_2$$\approx$0,4), slightly triaxial, minimum 
 becomes yrast at $I$$\approx$50.

The nucleus $^{108}$Ru (Fig.~\ref{TRS108Ru}) 
is representative of  a $\gamma$-soft 
system. Here, the competition between 
several near-yrast triaxial minima produces
 dramatic shape changes.
Although the triaxial minima are rather shallow at
$\omega$=0 (see Fig.~\ref{PESRub}), they become  stabilized at
high spins due to quasi-particle alignment.
The superdeformed minimum with  $\beta_2$$\approx$0.5
appears at exceptionally  low spins. At $\hbar\omega$=0.6\,MeV, this minimum
lies only $\approx$2 MeV above the  triaxial yrast line.

\subsection{Shape Changes at High Spin}\label{HSdef}

The systematics of 
calculated equilibrium $\beta_2$ and $\gamma$  deformations 
for even-even nuclei with $N$=58-66 
are displayed in Figs.~\ref{BCSTRA-Sr} (Sr), 
\ref{BCSTRA-Zr} (Zr),
\ref{BCSTRA-Mo} (Mo), and
\ref{BCSTRA-Ru} (Ru). All the local minima that appear
 in calculated
TRS's are visualized  as 
 $I$-dependent trajectories in the ($\beta_2, \gamma$) 
 space. In the following discussion,
 rotational bands and their underlying intrinsic configurations
are grouped into several categories,
 according to their deformations and intruder content.
In most cases, the dependence of
equilibrium deformations on configuration and
angular momentum  
can be simply explained in terms of
two factors: (i)
the deformation softness of the PES at $I$=0; and (ii)
the position of the Fermi level, $\lambda$,
 relative to  an intruder high-$j$ shell.
In this context,  recall that
  the routhian of the
strongest-aligned quasi-particle has a minimum 
as a function of $\gamma$
that is strongly dependent on the position of $\lambda$ 
 within the shell. 
Namely, if the Fermi level lies at the bottom of the shell,
positive values of $\gamma$ are favored, for a half-filled shell 
$\gamma$$\approx$--30$^\circ$ is preferred, and for 
$\lambda$ in the upper half of the shell, the
aligned high-$j$ particle drives the core towards
$\gamma$$<$--60$^\circ$
\cite{[Fra83],[Che83a],[Ham83],[Ben84a]}. Of course,
the total shape polarization exerted by aligned
quasi-particles results from an
interplay between  the contributions of protons and neutrons.

\subsubsection{Prolate shapes with $\beta_2 \approx$ 0.27-0.35.}
 The well-deformed prolate shapes
 are typical of ground-state  rotational bands in Sr and Zr nuclei. 
As seen in Table \ref{table_prolate-oblate}, ground-state  deformations 
gradually
increase from  $\beta_2$$\approx$0.28 in $N$=58 isotones to
  $\beta_2$$\approx$ 0.33-35 in the heavier systems.   

For all Sr isotopes except $^{104}$Sr ($N$=66), the deformation of the 
aligned {(\nh)$^2$} 2-q.p.
configuration is nearly the same as that of the ground state band.
Indeed, the calculated values of 
$\beta_2$ are only slightly reduced  and the values of   $|\gamma|$
are close to zero.  
For $^{104}$Sr, the neutron Fermi level
is  higher in the {$1h_{11/2}$} shell, i.e.,
between [532 5/2] and [523 7/2] orbitals
(Fig. \ref{WSLevel}). Consequently
the {(\nh)} alignment is reduced  and the 
 {\nh}  
crossing is predicted at a
rotational frequency close to that of
the {(\pg)} crossing.

The  values of $\gamma$  calculated for
the  {(\nh)$^2$(\pg)$^2$} 4-q.p.
configurations   
depend on $N$ (see Fig.~\ref{BCSTRA-Sr}). 
This can be explained in terms of the polarizing
effect of the  aligned high-$j$ particles.
Indeed, for $N$=58 (bottom of the {\nh} shell), 
 $\gamma$ deformation of a 4-q.p. band is calculated to 
be positive, and  with increasing $N$ the system is driven to
negative $\gamma$-values.
Interestingly, this polarizing effect is not observed for  the 
{\hb} bands.
This can be explained in terms of a rather $\gamma$-rigid potential
of the vacuum  configuration (g-band).
 The  $\gamma$-driving force exerted
by the aligned {\hb} pair  is not sufficient to polarize
the system towards $\gamma$$\ne$0. 
However, the additional deformation-driving effect of
the  aligned {$1g_{9/2}$} protons is sufficient to make the system
triaxial.
As expected, the alignment of the second pair of {$1h_{11/2}$} neutrons
gives rise to a reduction of  $\beta_2$ in the  {\hhg} 6-q.p. 
configuration.
The exception  to this pattern is $^{96}$Sr
where the {\em increase} in $\beta_2$ can be attributed to
 the {\nb} alignment.

The Zr isotopes are $\gamma$-softer than the Sr isotopes
(cf. Figs.~\ref{TRS100Sr} and \ref{TRS102Zr}), and shape changes
at high spins are more pronounced.
The g-bands are predicted to be prolate. The deformation change is 
associated
with the {$1h_{11/2}$} neutron alignment. Indeed, as seen in 
Fig.~\ref{Rou102Zr}, displaying 
the quasi-particle routhians characteristic of
a well-deformed prolate g-band in $^{102}$Zr,
the proton {$1g_{9/2}$} 
crossing frequency is higher than the neutron 
$1h_{11/2}$ crossing frequency.
In most cases, the aligned  {\hb} bands are  triaxial, with 
slightly negative $\gamma$ values.    

For some  Sr and Zr isotopes, at large spins ($I$$\sim$30--50)
there appear 
prolate well-deformed minima with  $\beta_2$$\sim$0.32--0.38.
 The associated  configurations are  {\hhg}
and {\hhig}
(see  Fig.~\ref{Rou102Zr}).

The  experimental and calculated angular 
momentum alignment  for the
well-deformed near-prolate
yrast bands of $^{100}$Zr and  $^{102}$Zr
are presented in  Fig.~\ref{IWNormal}.
Although our calculations are adiabatic and hence unable to
reproduce the alignment pattern in the 
band crossing region, it is clearly seen that  
the first band crossing is associated with 
the {\nh} alignment.

\subsubsection{Triaxial  or near-oblate collective
minima with 
--75$^\circ$$\protect\lesssim$$\gamma$$\protect\lesssim$$-$15$^\circ$}
\label{triaxial_minima}

In the heavy
{\em Sr isotopes} there appear
excited minima at oblate shapes which give rise to 
rotational bands (g$'$-bands) at $\gamma$=--60$^\circ$.
 Except for the  $N$=58
isotope ($\beta_2$=0.17), associated $\beta_2$
deformations  are typically $\approx$0.25 and increase slightly with 
increasing $I$.
The oblate bands are predicted to lie at $E^*$$\gtrsim$1\,MeV above
the  yrast configuration.
At higher spins, $I$$\gtrsim$20,  triaxial bands
built upon  several aligned quasi-particles
[{\hg}, {\hhg}, {\hhig}] emerge.
Within these bands, $\beta_2$ values
vary
 smoothly, from $\beta_2$$\approx$0.3 to $\approx$0.22. 
The triaxial many-quasi-particle
bands extend over $I$$\approx$24-46 for $N$=60 and 62, $I$$\approx$18-52
for $N$=64, and $I$$\approx$30-52 for $N$=66.
They are highly excited ($E^*$$\gtrsim$2\,MeV above yrast).

The heavy {\em  Zr isotopes} with  $N$=58--62
are predicted to have 
 oblate collective bands  similar 
to those calculated for the  Sr isotopes.
For $N$=64 and 66, triaxial minima with negative $\gamma$ 
deformations are predicted. The
corresponding   configurations are {\hg} and {\hhg},
extending  from $I$$\approx$20 to $I$$\approx$42.
 In  $^{106}$Zr, a 8-q.p. {\hhig} band
is predicted at  high spins, $I$$\sim$50. However,
due to the large 
quasi-particle content, its collectivity is weak 
($\beta_2$$\approx$0.15, $\gamma$$\approx$--80$^\circ$).
The excitation energy of this band is  1-4 MeV above
the superdeformed (SD) yrast band.

The heavy {\em Mo isotopes} are predicted to have
triaxial g-bands with negative  $\gamma$ values.
For $N$=58 and 60, the triaxial minima are very shallow. 
The large
excursions in the  $\gamma$ direction shown  in 
Fig.~\ref{BCSTRA-Mo} are due to this 
softness (or $\gamma$-instability) and hence should not be taken literally.
The heavier Mo isotopes have  better
localized triaxial minima 
($\beta_2$$\sim$0.3, $\gamma$$\sim$$-$20$^\circ$).
 Consequently, the deformation changes in their g-bands are small.
 The quasi-particle routhian diagram, characteristic of
triaxial configurations in Mo and Ru isotopes, is presented  in
Fig.~\ref{Rou108Ru}. 
Figure~\ref{IWTriaxial} shows
the  experimental and calculated angular 
momentum alignment for the yrast band of $^{108}$Ru.
In contrast to the situation for prolate shape 
(Fig.~\ref{Rou102Zr}), here  the  {\pg}  alignment
is expected to occur at a lower frequency. 
Because of the deformation softness, shape changes
 with quasi-particle alignment are more dramatic in 
the Mo isotopes 
than in their  (more rigid)
 Sr and Zr isotones. In most cases, highly aligned bands
have $\gamma$$\lesssim$$-$30$^\circ$, and  their collectivity 
gradually decreases with $I$. For example,  
the {\hhig} and {\hhigg} configurations calculated at $I$$\approx$40-50,
have weakly deformed shapes with $\beta_2$$\approx$0.12.

In recent experimental work~\cite{[Smi96]}, 
transition quadrupole moments of high-spin states in the
neutron-rich  Mo isotopes
have been measured using the Doppler-profile method.
The observed reduction of quadrupole
moments at $I$$\approx$10 
has been interpreted in terms of 
a deformation change from  axial shapes towards
triaxial shapes   with 
$\gamma$$>$0. 
Our calculations do not confirm this suggestion. 
According to the results shown in Fig.~\ref{BCSTRA-Mo}, 
 deformations 
of triaxial g-bands
in, e.g., $^{104,106,108}$Mo are rather stable, and 
it is only after the {\nh} alignment that some small
reduction in $\beta_2$ is predicted. In our opinion, the decrease
in the transition quadrupole moments seen  experimentally
 is due to the
mixing between a g-band and 
a {\hb} band at $I$$\approx$10. Indeed, as noted in 
Ref.~\cite{[Smi96]}, the observed moments of inertia reveal a gradual increase
for $\hbar\omega$=0.2--0.4\,MeV, reflecting a  large-interaction
band crossing.
There is a 
 difference between the  TRS calculations presented in
Ref.~\cite{[Smi96]} and these in our study. Namely, the
previous study predicts
  positive
$\gamma$ deformation for the aligned {\hb}  band in $^{102}$Mo.
In our opinion, this discrepancy is due to the  different spin-orbit
parameterizations employed in these two  papers.
As discussed in Sec.~\ref{MODEL},  the spin-orbit
potential used in our study was optimized to reproduce the experimental
shape transition near  $N$=56. The change in the spin-orbit
strength implies a different position of the {$1h_{11/2}$} neutron shell and,
consequently, gives rise to a                                    
different predicted deformation pattern.

The neutron-rich
{\em Ru isotopes} have  $\gamma$-soft ground states which are stabilized
by rotation (see Fig.~\ref{TRS108Ru}).
 For $N$=60-66,  the g-bands correspond to triaxial
shapes with negative values of $\gamma$.
(The positive-$\gamma$ minima never are predicted to be
 yrast at low and medium spins.)
The deformation changes  in $^{106}$Ru 
due to {\nh} and {\pg} alignments
are predicted to be rather small, with
the 
g-band, {\hb} band, and {\hg} band having similar deformations,
($\beta_2,\gamma$)$\approx$(0.25, $-$20$^\circ$). 
The strongly
triaxial {\hhg} configuration in $^{106}$Ru becomes yrast at
$I$$\approx$40, but it is quickly crossed
 by a SD band.
The nuclei  $^{108,110}$Ru  
behave very similarly.
Their g-bands correspond to
 deformation ($\beta_2,\gamma$)$\approx$(0.28,$-$20$^\circ$).
The {\nh} alignment,  the {\pg} alignment, and the second {\nh} alignment 
produce  triaxial shapes with $\beta_2$$\approx$0.2, 
$\gamma$$\approx$$-$45$^\circ$.
The {\nh} crossing is  calculated to occur at 
$\hbar\omega$$\approx$0.35\,MeV for both nuclei. This
is slightly lower than the observed crossing  frequency, 
$\hbar\omega_{\rm exp}$$\approx$0.4\,MeV \cite{[Lu95]}
(see Figs.~\ref{IWTriaxial} and
\ref{Rou108Ru}).  

Properties of heavy Mo and Ru isotopes are often discussed
in terms of triaxial degrees of freedom. Experimentally,
the  $2_2^+$ states
lie  at fairly  low excitation energies
 ($\approx$700-800\,keV), indicating the importance
of $\gamma$ deformation.
For $^{108,110}$Ru, the $B(E2)$ branching ratios measured in  
the $\gamma$-band at low spin 
are   consistent with that of a  rigid triaxial rotor 
\cite{[Sha94a]}. 
However, it has also been  pointed out
that at higher spins the analysis with the rotational-vibrational
model  gives
a better agreement \cite{[Lu95]}. More recent calculations
based on the 
generalized collective model \cite{[Tro96]} suggest
 that the experimental
result is consistent with the assumption of a $\gamma$-soft potential
with a triaxial minimum.
It should also be pointed out  that the CHFB calculations with the 
pairing-plus-quadrupole Hamiltonian \cite{[Dev96],[Dev96a]}
yield
 triaxial shapes, and that
in the interacting boson model  analysis
of Ref.~\cite{[Gia95]} these nuclei are  calculated
to be near  the O(6) limit
rather than the SU(3) limit, thus
indicating  $\gamma$ softness.

 \subsubsection{Triaxial minima with $\beta_2$=0.20--0.35, 
15$^\circ$$\protect\lesssim$$\gamma$$\protect\lesssim$45$^\circ$}

Triaxial minima with $\gamma$$>$0 are predicted
at low angular momentum in the  Mo and Ru isotopes.
At $I$=0 these minima are intrinsically equivalent 
to the triaxial ground-state configurations with $\gamma$$<$0.
However, due to the large moment of inertia at $\gamma$$<$0,
the latter bands are usually favored by rotation.
In most cases, the  $\gamma$$>$0  bands have 
($\beta_2, \gamma$)$\approx$ (0.3, 15$^\circ$), and
the deformation decreases to ($\beta_2, \gamma$)
$\approx$(0.2, 30$^\circ$) in the corresponding
aligned  configurations. 
An exception is the nucleus
$^{100}$Mo, whose yrast line has ($\beta_2, \gamma$) $\approx$
(0.2, 30$^\circ$) up to $I$$\approx$25 (the
 negative-$\gamma$ minimum 
quickly disappears).

At intermediate
angular momentum ($I$$\approx$24--30, $\hbar\omega$$\approx$0.6--0.7\,MeV)
in $^{96,98,102}$Sr and $^{102}$Zr
 appear well-deformed minima  appear with
($\beta_2, \gamma$)$\approx$ (0.2, 30$^\circ$). 
These  correspond to the aligned  {\hhg} configuration 
and become yrast or near yrast at $I$$\approx$24.

At very high spins
($I$$\approx$50--70, $\hbar\omega$$\approx$1.0--1.5\,MeV)
 in $^{98,100,104,106}$Zr and $^{102,106,108}$Mo
strongly deformed triaxial minima with
 $\beta_2$=0.28--0.4 and $\gamma$$\approx$30$^\circ$ are predicted.
These bands involve the {\ph} and {\na} intruders
(see Figs.~\ref{BCSTRA-Zr} and \ref{BCSTRA-Mo}).

\subsubsection{Noncollective configurations}

Noncollective many-quasi-particle
states ($\gamma$=60$^\circ$)
correspond to the 
optimal (stretched) shell-model configurations with 
the single-particle angular momenta of 
all decoupled
nucleons aligned along the axis 
of the total angular momentum.
Often, the stretched states correspond to termination
points of collective rotational bands.
The many-quasi-particle bands with small $\beta_2$ deformations
and/or $\gamma$-values close to the $\gamma$=60{$^\circ$} limit
(seen in Figs.~\ref{BCSTRA-Sr}-\ref{BCSTRA-Ru}) are candidates
for  such terminating structures. (For
recent calculations of terminating
configurations in this mass region, see Ref.~\cite{[Rag96]}.)

In all nuclei considered, the TRS
calculations predict that  the noncollective configurations 
approach yrast in the spin range $I$=30--50.
The corresponding deformations
increase with increasing
 $I$, approaching   $\beta_2$$\approx$0.3--0.35 at 
$I$$\approx$60. Since the main focus of 
our paper is on collective rotation and shape effects, and since
the adiabatic
TRS method used does not allow for a detailed description of
optimal states, we do not discuss noncollective configurations
quantitatively.

\subsubsection{Superdeformed near-prolate  minima with 
 $\beta_2$$\protect\gtrsim$0.40} 
The nuclei from the heavy-Zr region are 
excellent  candidates for
finding  SD bands.
In our calculations, 
SD  minima have been predicted in  $^{102-110}$Ru, $^{100-108}$Mo,
$^{102-106}$Zr, and $^{100}$Sr. 
 These minima approach yrast at high rotational frequencies,
typically at $\hbar\omega$$\gtrsim$0.8\,MeV. Since 
at these high angular momenta  pairing correlations are less important
\cite{[Naz85]},
intrinsic configurations of SD states are
well characterized by the
intruder orbitals carrying
large  principal oscillator
numbers ${\cal N}$
(high-${\cal N}$ orbitals) \cite{[Naz89],[Ben88],[Jan91],[Bak95]}.
In this mass region, 
these are the  ${\cal N}$=5  proton and
 ${\cal N}$=6  neutron states.
 Because of their large intrinsic
angular momenta and quadrupole moments,
 high-${\cal N}$ orbitals strongly respond to the Coriolis
interaction and to the deformed average field. Consequently,
their occupation numbers are  good  indicators 
of rotation and deformation properties of SD bands.

{}Figure~\ref{RouSD} shows the  single-particle routhian
diagram representative of discussed SD configurations.
The single-particle gaps in the routhian spectrum 
indicate the best  candidates for  SD bands: at low spins,
the combination of the 
$Z$=42 and $N$=58 gaps leads to $^{100}$Mo, while
 the Mo, Tc, and Ru isotopes 
with $N$ approaching 68 are expected to favor SD high spin bands.
This observation is fully corroborated by our calculations.
Indeed, the {\nzpz} and {\ntpt} SD bands in
$^{100}$Mo and 
 $^{110}$Ru, respectively,  are predicted to appear
near yrast already at $I$$\approx$22.

For  the heavy Sr isotopes, SD minima 
are calculated only for
 $^{100}$Sr, and only at  very high spins,
$I$$\approx$64--70. The
SD band contains two ${\cal N}$=6 neutrons and one
${\cal N}$=5 proton; hence the
 corresponding configuration is labeled
{\ntpo}. This configuration is also predicted 
 in
 $^{102}$Zr and $^{104}$Zr. The corresponding 
SD bands behave rather smoothly with $\omega$
 up to the highest spins considered.
In contrast,
for $^{106}$Zr the SD yrast line is less
regular. Here, the 
{\nopz} band
is crossed at $\hbar\omega$$\approx$1.1\,MeV
by the more strongly deformed {\ntpo} band.
At still higher spins, there is
a transition to an  even more deformed structure, {\nrpt}.
The corresponding kinematic moments of inertia, 
{${\cal J}^{(1)}$},
extracted from 
the TRS calculations are displayed in Fig.~\ref{IWSuper}.
It is seen that the moments of inertia in  all  {\ntpo} bands
decrease gradually with frequency. The increased 
{${\cal J}^{(1)}$} in  $^{106}$Zr is due to the two band crossings
discussed above.

At low spin,
the SD minimum of the  $N$=58 isotones of
Mo and Ru ($^{100}$Mo and $^{102}$Ru)
can be associated with  the configuration {\nzpz}.
At $\hbar\omega$$\approx$1\,MeV, 
the configuration {\nopt} becomes lower in energy
(see Fig.~\ref{IWSuper}).
It is  noted that 
$^{100}$Mo 
is among the few nuclei,
together with $^{108}$Mo and $^{110}$Ru, in which the
SD
minimum appears at particularly
 low spins, i.e.,  at  $I$$\approx$20 (the corresponding SD band
is calculated  to appear 
$\approx$3\,MeV above yrast). As discussed above,
the microscopic reason 
for this structure can be explained
in terms
of   the SD shell gaps
at $Z$=42 and $N$=58.
The nuclei $^{102,104,106}$Mo
and $^{106,108}$Ru are predicted to have rather
regular SD bands associated with  the 
  {\ntpo} and {\ntpt} 
configurations. On the other hand, the SD yrast line of
 $^{108}$Mo and $^{104}$Ru is expected to be less regular, with several
consecutive band crossings.
The best candidates for 
observing  high-spin SD bands in this mass region are
$^{110,112}$Ru.

\section{Conclusions}\label{CONCLUSIONS}

The macroscopic-microscopic method based on the deformed Woods-Saxon
potential
was applied to neutron-rich nuclei from the 
$A$$\approx$100 mass region.  
Thanks to a number of experimental developments, one is now able
to obtain  spectroscopic information on
 these systems,
and more data should become
  be available  in the near future, due to the large gamma-ray 
arrays such as GammaSphere or EuroBall, and the advent of
neutron-rich radioactive beams.

The main focus of this study is on the deformation properties 
and shape coexistence effects at low and high angular momentum.
High-spin calculations were carried out for
the vacuum configurations (i.e., 
 positive-parity, even-spin)
in
even-even Sr, Zr, Mo, and Ru isotopes with
58$\leq$$N$$\leq$66.  

The single-particle model used was previously optimized
to reproduce the  transition from spherical  to deformed
shapes at $N$$\approx$58. The calculated ground-state deformations
for heavier systems are
in general  agreement with experimental data.

The influence of particle number projection has been investigated. In general, 
the potential energy surfaces calculated with the PNP method 
have better developed deformed minima.
A replacement of the energies at spin zero,
$ E^{\omega=0}$ in Eq.~(\ref{e101}), calculated within the  BCS by 
 those calculated with PNP,  has only  minor consequences for
predicted high-spin properties. 

At low and medium spins, the alignment pattern is governed by an
alignment of the 
{\nh} and {\pg} quasi-particles. The associated deformation changes
are explained in terms of the shape polarization exerted by aligned high-$j$
quasi-particles. According to the calculations, the first band crossing
in the neutron-rich nuclei considered can be associated with the
alignment of the $1h_{11/2}$ neutron pair. The frequency of
the second bandcrossing, due to the
{$1g_{9/2}$} proton alignment, is rather  high in the 
well-deformed Sr and 
Zr isotopes, but it becomes comparable with the neutron crossing frequency
in the triaxial Mo and Ru isotopes.

At higher angular momenta,  the high-$\cal N$ intruder orbitals, namely
$\cal N$=5 protons and $\cal N$=6 neutrons,
become important. They are 
occupied  in superdeformed structures and in some
highly-triaxial configurations.

{}From the point of view of nuclear quadrupole collectivity, 
probably the most interesting
nuclei in the neutron-rich $A$$\approx$100 mass region are systems 
near
$^{104}$Mo and $^{106}$Ru, which are
predicted to have stable triaxial shapes
($\gamma$$\approx$--30{$^\circ$}) in 0-q.p., 2-q.p., and 4-q.p. configurations.
Although the question of whether they are $\gamma$-soft or $\gamma$-deformed
at low spins has not yet been settled, 
these systems seem to form a very good testing ground
for theoretical models of nuclear triaxiality. In particular at higher spins, 
where the triaxial minima are predicted to be deeper, the shape with
$\gamma$$\approx$--30{$^\circ$} can give rise to interesting selection rules
associated with the effective $C_4$ symmetry of the Hamiltonian \cite{[Ham87a]}.
Finally, the most favorable candidates for superdeformation in this mass region
are $^{100}$Mo and $^{110,112}$Ru in which 
 the SD minima are predicted
at particularly low spins.

\acknowledgments
 One of the authors (J.S.) would like to express his thanks to
 Bill Phillips and the staff of the Department of Nuclear Physics 
 of Manchester University 
 for the hospitality extended to him during his stay there.
 The helpful assistance of Mike Pettipher of 
 the Manchester Computational Center is gratefully acknowledged. 
The authors would also like to thank Jerry Garrett and Bill Walters 
for useful comments.
Oak Ridge National
Laboratory is managed for the U.S. Department of Energy
 by Lockheed Martin Energy Research
Corporation  under Contract No.
DE-AC05-96OR22464.  
The Joint Institute for Heavy Ion
 Research has as member institutions the University of Tennessee,
Vanderbilt University, and the Oak Ridge National Laboratory; it
is supported by the members and by the Department of Energy
through Contract No. DE-FG05-87ER40361 with the University
of Tennessee.  Theoretical nuclear physics research 
at the University of Tennessee  is supported by the U.S. Department of
Energy through Contract
DE-FG02-96ER40963.
 This work has been partially supported by 
 the Polish State Committee for Scientific Research
under Contracts No. 20450~91~01 and
20954~91~01.  


\begin{thebibliography}{100}

\bibitem{[Zir88]}
{{\it Nuclear Structure of the Zirconium Region}, eds. J. Eberth, R.A. Meyer
  and K. Sistemich (Springer-Verlag, 1988)}.

\bibitem{[Woo92]}
{J.L. Wood, K. Heyde, W. Nazarewicz, M. Huyse and P. van Duppen, Phys. Rep.
  {\bf 215}, 101 (1992)}.

\bibitem{[Lhe94]}
{G. Lhersonneau, B. Pfeiffer, K.-L. Kratz, T. Enqvist, P.P. Jauho, A. Jokinen,
  J. Kantele, M. Leino, J.M. Parmonen, H. Penttil\"a, J. \"Ayst\"o and the
  ISOLDE collaboration, Phys. Rev. {\bf C49}, 1379 (1994)}.

\bibitem{[Ber94]}
{M. Bernas, S. Czajkowski, P. Armbruster, H. Geissel, Ph. Dessagne, C. Donzaud,
  H-R. Faust, E. Hanelt, A. Heinz, M. Hesse, C. Kozhuharov, Ch. Miehe, G.
  M{\"u}nzenberg, M. Pf{\"u}tzner, C. R{\"o}hl, K.-H. Schmidt, W. Schwab, C.
  St{\'e}phan, K. S{\"u}mmerer, L. Tassan-Got and B. Voss, Phys. Lett. {\bf
  B331}, 19 (1994)}.

\bibitem{[Ays92]}
{J. \"Ayst\"o, A. Astier, T. Enqvist, K. Eskola, Z. Janas, A. Jokinen, K.-L.
  Kratz, M. Leino, H. Penttil\"a, B. Pfeiffer and J. \.Zylicz, Phys. Rev.
  Lett. {\bf 69}, 1167 (1992)}.

\bibitem{[Meh96]}
{T. Mehren, B. Pfeiffer, S. Schoedder, K.-L. Kratz, M. Huhta, P. Dendooven, A.
  Honkanen, G. Lhersonneau, M. Oinonen, J.-M. Parmonen, H. Penttil{\"a}, A.
  Popov, V. Rubchenya and J. {\"A}yst{\"o}, Phys. Rev. Lett. {\bf 77}, 458
  (1996)}.

\bibitem{[Hot90]}
{M.A.C. Hotchkis, J.L. Durell, J.B. Fitzgerald, A.S. Mowbray, W.R. Phillips, I.
  Ahmad, M.P. Carpenter, R.V.F. Janssens, T.L. Khoo, E.F. Moore, L.R. Morss,
  Ph. Benet and D. Ye, Phys. Rev. Lett. {\bf 64}, 3123 (1990)}.

\bibitem{[Hot91]}
{M.A.C. Hotchkis, J.L. Durell, J.B. Fitzgerald, A.S. Mowbray, W.R. Phillips, I.
  Ahmad, M.P. Carpenter, R.V.F. Janssens, T.L. Khoo, E.F. Moore, L.R. Morss,
  Ph. Benet and D. Ye, Nucl. Phys. {\bf A530}, 111 (1991)}.

\bibitem{[Lu95]}
{Q.H. Lu, K. Butler-Moore, S.J. Zhu, J.H. Hamilton, A.V. Ramayya, V.E.
  Oberacker, W.C. Ma, B.R.S. Babu, J.K. Deng, J. Kormicki, J.D. Cole, R.
  Aryaeinejad, Y.X. Dardenne, M. Drigert, L.K. Peker, J.O. Rasmussen, M.A.
  Stoyer, S.Y. Chu, K.E. Gregorich, I.Y. Lee, M.F. Mohar, J.M. Nitschke, N.R.
  Johnson, F.K. McGowan, G.M. Ter-Akopian, Yu.Ts. Oganessian and J.B. Gupta,
  Phys. Rev. {\bf C52}, 1348 (1995)}.

\bibitem{[Hae86]}
{D.R. Haenni, H. Dejbaksh, R.P. Schmitt and G. Mouchaty, Phys. Rev. {\bf C33},
  1543 (1986)}.

\bibitem{[Est89]}
{R. Estep, R.K. Sheline, D.J. Decman, E.A. Henry, L.G. Mann, R.A. Mayer, W.
  Stoeff, L.E. Ussery and J. Kantele, Phys. Rev. {\bf C39}, 76 (1989)}.

\bibitem{[Gel88]}
{W. Gelletly, Y. Abdelrahman, A.A. Chishti, J.L. Durell, J. Fitzgerald, C.J.
  Lister, J.H. McNeill, W.R. Phillips and B.J. Varley, in {\it Nuclear
  Structure of the Zirconium Region}, ed. by J. Eberth, R.A. Meyer and K.
  Sistemich, (Springer-Verlag, 1988), p. 102}.

\bibitem{[Dej95]}
{H. Dejbakhsh and S. Bouttchenko, Phys. Rev. {\bf C52}, 1810 (1995)}.

\bibitem{[Sha94a]}
{J.A. Shannon, W.R. Phillips, J.L. Durell, B.J. Varley, W. Urban, C.J. Pearson,
  I. Ahmad, C.J. Lister, L.R. Morss, K.L. Nash, C.W. Williams, N. Schulz, E.
  Lubkiewicz and M. Bentaleb, Phys. Lett. {\bf 336B}, 136 (1994)}.

\bibitem{[Sch95a]}
{S. Schoedder, G. Lhersonneau, A. Wohr, G. Skarnemark, J. Alstad, A. Nahler, K.
  Eberhardt, J. Aysto, N. Trautmann and K.-L.Kratz, Z. Phys. {\bf A352}, 237
  (1995)}.

\bibitem{[Reg96]}
{P.H. Regan, T.P.S.M. Menezes, C.J. Pearson, W. Gelletly, C.S. Purry, P.
  Walker, O. Burglin, S. Juutinen, R. Julin, K. Helariutta, A. Savelius, P.
  Jones, P.A. Butler, G. Jones, P. Greenless and R. Wyss,
 to be published}.

\bibitem{[Men96]}
{T.P.S.M. Menezes, P.H. Regan, P.M. Walker, W. Gelletly, C.J. Pearson, C.S.
  Purry, R. Julin, S. Juutinen, P. Jones, A. Savelius, K. Helariutta, H.
  Kankaap{\"o}, P.A. Butler, G. Jones and P. Greenless, Proc. Conf. on Nuclear
  Structure at the Limits, Argonne 1996}.

\bibitem{[Dur95]}
{J.L. Durell, W.R. Phillips, C.J. Pearson, J.A. Shannon, W. Urban, B.J. Varley,
  N. Rowley, K. Jain, I. Ahmad, C.J. Lister, L.R. Morss, K.L. Nash, C.W.
  Williams, N. Schulz, E. Lubkiewicz and M. Bentaleb, Phys. Rev. {\bf C52},
  R2306 (1995)}.

\bibitem{[Dur96]}
{J.L. Durell, W.R. Phillips, C.J. Pearson, J.A. Shannon, W. Urban, B.J. Varley,
  I. Ahmad, C.J. Lister, L.R. Morss, K.L. Nash, C.W. Williams, N. Schulz, E.
  Lubkiewicz and M. Bentaleb, Proc. Conf. on Nuclear Structure at the Limits,
  Argonne 1996}.

\bibitem{[Smi96]}
{A.G. Smith, J.L. Durell, W.R. Phillips, M.A. Jones, M. Leddy, W. Urban, B.J.
  Varley, I. Ahmad, L.R. Morss, M. Bentaleb, A. Guessous, E. Lubkiewicz, N.
  Schulz and R. Wyss, Phys. Rev. Lett. {\bf 77}, 1711 (1996)}.

\bibitem{[Ays90]}
{J. \"Ayst\"o, P.P. Jauho, Z. Janas, A. Jokinen, J.M. Parmonen, H. Penttila, P.
  Taskinen, R. Beraud, R. Duffait, A. Emsallem, J. Meyer, M. Meyer, N. Redon,
  M.E. Leino, K. Eskola and P. Dendooven, Nucl. Phys. {\bf A515}, 365 (1990)}.

\bibitem{[Gue95]}
{A. Guessous, N. Schulz, W.R. Phillips, I. Ahmad, M. Bentaleb, J.L. Durell,
  M.A. Jones, M. Leddy, E. Lubkiewicz, L.R. Morss, R. Piepenbring, A.G. Smith,
  W. Urban and B.J. Varley, Phys. Rev. Lett. {\bf 75}, 2280 (1995)}.

\bibitem{[Gue96]}
{A. Guessous, N. Schulz, M. Bentaleb, E. Lubkiewicz, J.L. Durell, C.J. Pearson,
  W.R. Phillips, J.A. Shannon, W. Urban, B.J. Varley, I. Ahmad, C.J. Lister,
  L.R. Morss, K.L. Nash, C.W. Williams and S. Khazrouni, Phys. Rev. {\bf C53},
  1191 (1996)}.

\bibitem{[Lhe95b]}
{G. Lhersonneau, H. Gabelmann, B. Pfeiffer, K.-L. Kratz and the ISOLDE
  Collaboration, Z. Phys. {\bf A352}, 293 (1995)}.

\bibitem{[Lhe95]}
{G. Lhersonneau, H. Gabelmann, M. Liang, B. Pfeiffer, K.-L. Kratz, H. Ohm and
  the ISOLDE Collaboration, Phys. Rev. {\bf C51}, 1211 (1995)}.

\bibitem{[Lhe96]}
{G. Lhersonneau, P. Dendooven, A. Honkanen, M. Muhta, M. Oinonen, P.
  Penttil{\"a}, J. {\"A}yst{\"o}, J. Kurpeta, J.R. Persson and A. Popov, Phys.
  Rev. {\bf C54}, 1592 (1996)}.

\bibitem{[Ohm91]}
{H. Ohm, M. Liang, U. Paffrath, B. De Sutter, K. Sistemich, A.-M. Schmitt, N.
  Kaffrell, N. Trautmann, T. Seo, K. Shizuma, G. Moln\'ar, K. Kawade and R.A.
  Meyer, Z. Phys. {\bf A340}, 5 (1991)}.

\bibitem{[Lia93]}
{M. Liang, H. Ohm, B. De Sutter and K. Sistemich, Z. Phys. {\bf A344}, 357
  (1993)}.

\bibitem{[Lia93a]}
{M. Liang, H. Ohm, I. Ragnarsson and K. Sistemich, Z. Phys. {\bf A346}, 101
  (1993)}.

\bibitem{[Lia95]}
{M. Liang, H. Ohm, B. De Sutter-Pomm{\'e} and K. Sistemich, Z. Phys. {\bf
  A351}, 13 (1995)}.

\bibitem{[Dej88]}
{H. Dejbakhsh, R.P. Schmitt and G. Mouchaty, Phys. Rev. {\bf C37}, 621 (1988)}.

\bibitem{[But95]}
{K. Butler-Moore, R. Aryaeinejad, J.D. Cole, Y. Dardenne, P.G. Greenwood, J.H.
  Hamilton, A.V. Ramayya, W.-C. Ma, B.R.S. Babu, J.O. Rasmussen, M.A. Stoyer,
  S.Y. Chu, K.E. Gregorich, M. Mohr, S. Asztalus, S.G. Prussin, M.J. Moody,
  R.W. Lougheed and J.F. Wild, Phys. Rev. {\bf C52}, 1339 (1995)}.

\bibitem{[Pfe95a]}
{B. Pfeiffer, G. Lhersonneau, H. Gabelmann, K.-L. Kratz and the ISOLDE
  Collaboration, Z. Phys. {\bf A353}, 1 (1995)}.

\bibitem{[Lhe95a]}
{G. Lhersonneau, B. Pfeiffer, M. Huhta, A. W{\"o}hr, L. Kl{\"o}ckl, K.-L.
  Kratz, J. {\"A}yst{\"o}, The ISOLDE Collaboration, Z. Phys. {\bf A351}, 357
  (1995)}.

\bibitem{[Ben84]}
{R. Bengtsson, P. M\"oller, J.R. Nix and J.-y. Zhang, Phys. Scr. {\bf 29}, 402
  (1984)}.

\bibitem{[Naz88]}
{W. Nazarewicz and T. Werner, in {\it Nuclear Structure of the Zirconium
  Region}, ed. by J. Eberth, R.A. Meyer and K. Sistemich, (Springer-Verlag,
  1988), p. 277}.

\bibitem{[Rag84]}
{I. Ragnarsson and R.K. Sheline, Phys. Scr. {\bf 29}, 385 (1984)}.

\bibitem{[Rag88]}
{I. Ragnarsson and T. Bengtsson, in {\it Nuclear Structure of the Zirconium
  Region}, eds. J. Eberth, R.A. Meyer and K. Sistemich (Springer-Verlag, 1988),
  p. 193}.

\bibitem{[Naz88a]}
{W. Nazarewicz, in {\sl Contemporary Topics in Nuclear Structure Physics} eds.
  R.F. Casten, A. Frank, M. Moshinsky and S. Pittel (World Scientific,
  Singapore, 1988) 467}.

\bibitem{[Dob88]}
{J. Dobaczewski, W. Nazarewicz, J. Skalski and T.R. Werner, Phys. Rev. Lett.
  {\bf 60}, 2254 (1988)}.

\bibitem{[Dob88a]}
{J. Dobaczewski, in {\sl Contemporary Topics in Nuclear Structure Physics} eds.
  R.F. Casten, A. Frank, M. Moshinsky and S. Pittel (World Scientific,
  Singapore, 1988) 227}.

\bibitem{[Wer94]}
{T.R. Werner, J. Dobaczewski, M.W. Guidry, W. Nazarewicz and J.A. Sheikh,
  Nucl. Phys. {\bf A578}, 1 (1994)}.

\bibitem{[Ars69]}
{D.A. Arseniev, A. Sobiczewski and V.V. Soloviev, Nucl. Phys. {\bf A139}, 269
  (1969)}.

\bibitem{[She72]}
{R.K. Sheline, I. Ragnarsson and S.G. Nilsson, Phys. Lett. {\bf 41B}, 115
  (1972)}.

\bibitem{[Fae74]}
{A. Faessler, J.E. Galonska, U. G\"otz and H.C. Pauli, Nucl. Phys. {\bf A230},
  302 (1974)}.

\bibitem{[Fed78]}
{P. Federman, S. Pittel and R. Campos, Phys. Lett. {\bf 82B}, 9 (1978)}.

\bibitem{[Fed78a]}
{P. Federman and S. Pittel, Phys. Lett. {\bf 77B}, 29 (1978)}.

\bibitem{[Azu79]}
{ R.E. Azuma, G.L. Borchert, L.C. Carraz, P.G. Hansen, B. Jonson, S. Mattsson,
  O.B. Nielsen, G. Nyman, I. Ragnarsson and H.L. Ravn, Phys. Lett. {\bf 86B}, 5
  (1979)}.

\bibitem{[Cam80]}
{X. Campi and M. Epherre, Phys. Rev. {\bf C22}, 2605 (1980)}.

\bibitem{[Mol81a]}
{P. M\"oller and J.R. Nix, At. Data Nucl. Data Tables {\bf 26}, 165 (1981)}.

\bibitem{[Kho82a]}
{S.K. Khosa, P.N. Tripathi and S.K. Sharma, Phys. Lett. {\bf 119B}, 257
  (1982)}.

\bibitem{[Abe82]}
{S. {\AA}berg, Phys. Scr. {\bf 25}, 23 (1982)}.

\bibitem{[Hey84a]}
{K. Heyde, J. Moreau and M. Waroquier, Phys. Rev. {\bf C29}, 1859 (1984)}.

\bibitem{[Kum85a]}
{A. Kumar and M.R. Gunye, Phys. Rev. {\bf C32}, 2116 (1985)}.

\bibitem{[Bon85]}
{P. Bonche, H. Flocard, P.H. Heenen, S.J. Krieger and M.S. Weiss, Nucl. Phys.
  {\bf A443}, 39 (1985)}.

\bibitem{[Cas85c]}
{R.F. Casten, W. Frank and P. von Brentano, Nucl. Phys. {\bf A444}, 133
  (1985)}.

\bibitem{[Gal86]}
{D. Galeriu, D. Bucurescu and M.J. Ivascu, J. Phys. {\bf G12}, 329 (1986)}.

\bibitem{[Hey88]}
{K. Heyde, in {\it Nuclear Structure of the Zirconium Region}, ed. by J.
  Eberth, R.A. Meyer and K. Sistemich, (Springer-Verlag, 1988), p. 3}.

\bibitem{[Sha88a]}
{S.K. Sharma, P.N. Tripathi and S.K. Khosa, Phys. Rev. {\bf C38}, 2935 (1988)}.

\bibitem{[Sug90]}
{M. Sugita and A. Arima, Nucl. Phys. {\bf A515}, 77 (1990)}.

\bibitem{[Que90]}
{P. Quentin, N. Redon, J. Meyer and M. Meyer, Phys. Rev. {\bf C41}, 341
  (1990); Erratum Phys. Rev. {\bf C43}, 361 (1991)}.

\bibitem{[Tro91]}
{D. Troltenier, J.A. Maruhn, W. Greiner, V. Velazquez Aguilar, P.O. Hess and
  J.H. Hamilton, Z. Phys. {\bf A338}, 261 (1991)}.

\bibitem{[Dej92]}
{H. Dejbakhsh, D. Latypov, G. Ajupova and S. Schlomo, Phys. Rev. {\bf C46},
  2326 (1992)}.

\bibitem{[Cha91]}
{R.R. Chasman, Z. Phys. {\bf A339}, 111 (1991)}.

\bibitem{[Abo92]}
{Y. Aboussir, J.M. Pearson, A.K. Dutta and F. Tondeur, Nucl. Phys. {\bf A549},
  155 (1992); J.M. Pearson, private communication, 1993}.

\bibitem{[Kir93]}
{E. Kirchuk, P. Federman and S. Pittel, Phys. Rev. {\bf C47}, 567 (1993)}.

\bibitem{[She93]}
{J.A. Sheikh and P. Ring, Phys. Rev. {\bf C47}, R1850 (1993)}.

\bibitem{[Hir93]}
{D. Hirata, H. Toki, I. Tanihata and P. Ring, Phys. Lett. {\bf 314B}, 168
  (1993)}.

\bibitem{[Ska93]}
{J. Skalski, P.-H. Heenen and P. Bonche, Nucl. Phys. {\bf A559}, 221 (1993)}.

\bibitem{[Buc94]}
{F. Buchinger, J.E. Crawford, A.K. Dutta, J.M. Pearson and F. Tondeour, Phys.
  Rev. {\bf C49}, 1402 (1994)}.

\bibitem{[Bha94]}
{A. Bharti and S.K. Khosa, Nucl. Phys. {\bf A572}, 317 (1994)}.

\bibitem{[Bha94a]}
{A. Bharti, R. Devi and S.K. Khosa, J. Phys.{\bf G20}, 1231 (1994)}.

\bibitem{[Bar95a]}
{A. Baran and W. H{\"o}henberger, Phys. Rev. {\bf C52}, 2242(1995)}.

\bibitem{[Mol95]}
{P. M\"{o}ller, J.R. Nix, W.D. Myers and W.J. Swiatecki, Atom. Data and Nucl.
  Data Tables {\bf 59}, 185 (1995)}.

\bibitem{[Laz95]}
{G.A. Lalazissis, M.M. Sharma, Nucl. Phys. {\bf A586}, 201 (1995)}.

\bibitem{[Gia95]}
{A. Giannatiempo, A. Nannini, P. Sona and D. Cutoiu, Phys. Rev. {\bf C52}, 2969
  (1995)}.

\bibitem{[Lon95]}
{G.L. Long, S.J. Zhu, L. Tian and H.Z. Sun, Phys. Lett. {\bf B345}, 351 (1995)}.

\bibitem{[Hey95]}
{K. Heyde, C. De Coster, P. Van Isacker, J. Jolie and J.L. Wood, Phys. Scr.
  {\bf T56}, 133 (1995)}.

\bibitem{[Cos96]}
{C. De Coster, B. Decroix and K. Heyde, Phys. Lett. {\bf 379B}, 20 (1996)}.

\bibitem{[Cos96a]}
{C. De Coster, K. Heyde, B. Decroix, P. Van Isacker, J. Jolie, H. Lehmann and
  J. L. Wood, Nucl. Phys. {\bf A600}, 251 (1996)}.

\bibitem{[Sin96]}
{A.J. Singh and P.K Paina, Phys. Rev. {\bf C53}, 1228 (1996)}.

\bibitem{[Tro96]}
{D. Troltenier, J.P.Draayer, B.R.S. Babu, J.H. Hamilton, A.V. Ramayya and V.E.
  Oberacker, Nucl. Phys. {\bf A601}, 56 (1996)}.

\bibitem{[Dev96]}
{R. Devi and S.K. Khosa, Phys. Rev. {\bf C54}, 1661 (1996)}.

\bibitem{[Dev96a]}
{R. Devi, A. Pandoh and S.K. Khosa, Z. Phys. {\bf A355}, 389 (1996)}.

\bibitem{[Rag96]}
{I. Ragnarsson, A.V. Afanasjev and J. Gizon, Z. Phys.
{\bf A355}, 383 (1996)}.

\bibitem{[Gra76]}
{A. Grau, E.L. Samuelson, F.A. Rickey, P.C. Simms and G.J. Smith, Phys. Rev.
  {\bf C14}, 2297 (1976)}.

\bibitem{[Voi76]}
{M.A.J. deVoigt, J.F.W. Jansen, F. Bruining and Z. Sujkowski, Nucl. Phys. {\bf
  A270}, 141 (1976)}.

\bibitem{[Fra85]}
{S. Frauendorf, in {\sl Proc. Int. Symp. In-Beam Nuclear Spectroscopy},
  Debrecen (1984)}.

\bibitem{[Kel85]}
{H.-J. Keller, S. Frauendorf, U. Hagemann, L. K\"aubler, H. Prade and F. Stary,
  Nucl. Phys. {\bf A444}, 261 (1985)}.

\bibitem{[Kei95]}
{M. Keim, E. Arnold, W. Borchers, U. Georg, A. Klein, R. Neugart, L. Vermeeren,
  R.E. Silverans and P. Lievens, Nucl. Phys. {\bf A586}, 219 (1996)}.

\bibitem{[Lie96]}
{P. Lievens, E. Arnold, W. Borchers, U. Georg, M. Keim, A. Kein, R. Neugart, L.
  Vermeeren and R.E. Silverans, Europhys. Lett. {\bf 22}, 11 (1996)}.

\bibitem{[Str67]}
{V.M. Strutinsky, Nucl. Phys. {\bf A95}, 420 (1967)}.

\bibitem{[Bra72]}
{M. Brack, J. Damg\aa rd, A.S. Jensen, H.C. Pauli, V.M. Strutinsky and C.Y.
  Wong, Rev. Mod. Phys. {\bf 44}, 320 (1972)}.

\bibitem{[Cwi87]}
{S. \'Cwiok, J. Dudek, W. Nazarewicz, J. Skalski and T. Werner, Comput. Phys.
  Commun. {\bf 46}, 379 (1987)}.

\bibitem{[Mye69]}
{W.D. Myers and W.J. Swiatecki, Ann. Phys. (N.Y.) {\bf 55}, 395 (1969)}.

\bibitem{[Kra79]}
{H.J. Krappe, J.R. Nix and A.J. Sierk, Phys. Rev. {\bf C20}, 992 (1979)}.

\bibitem{[Mol81]}
{P. M\"oller and J.R. Nix, Nucl. Phys. {\bf A361}, 117 (1981)}.

\bibitem{[Naz84]}
{W. Nazarewicz, P. Olanders, I. Ragnarsson, J. Dudek, G.A. Leander, P. M\"oller
  and E. Ruchowska, Nucl. Phys. {\bf A429}, 269 (1984)}.

\bibitem{[Nil95]}
{S.G. Nilsson and I. Ragnarsson, {\em Shapes and Shells in Nuclear Structure}
  (Cambridge University Press, Cambridge 1995)}.

\bibitem{[Jon96]}
{L.-O. J\"onsson, Nucl. Phys. {\bf A608}, 1 (1996)}.

\bibitem{[Naz90b]}
{W. Nazarewicz, Nucl. Phys. {\bf A520}, 333c (1990)}.

\bibitem{[Naz90]}
{W. Nazarewicz, M.A. Riley and J.D. Garrett, Nucl. Phys. {\bf A512}, 61
  (1990)}.

\bibitem{[Mol92b]}
{P. M\"oller and J.R. Nix, Nucl. Phys. {\bf A536}, 20 (1992)}.

\bibitem{[Dud81]}
{J. Dudek, Z. Szyma\'nski and T. Werner, Phys. Rev. {\bf C23}, 920 (1981)}.

\bibitem{[Naz85]}
{W. Nazarewicz, J. Dudek, R. Bengtsson, T. Bengtsson and I. Ragnarsson, Nucl.
  Phys. {\bf A435}, 397 (1985)}.

\bibitem{[Naz87]}
{W. Nazarewicz, G.A. Leander and J. Dudek, Nucl. Phys. {\bf A467}, 437 (1987)}.

\bibitem{[Naz89]}
{W. Nazarewicz, R. Wyss and A. Johnson, Nucl. Phys. {\bf A503}, 285 (1989)}.

\bibitem{[Rei96]}
{P.-G. Reinhard, W. Nazarewicz, M. Bender and J.A. Maruhn, Phys. Rev. {\bf
  C53}, 2776 (1996)}.

\bibitem{[Ram87]}
{S. Raman, C.H. Malarkey, W.T. Milner, C.W. Nestor, Jr. and P.H. Stelson,
  Atomic Data Nucl. Data Tables {\bf 36}, 1 (1987)}.

\bibitem{[Rin82]}
{P. Ring, A. Hayashi, K. Hara, H. Emling and E. Grosse, Phys. Lett. {\bf 110B},
  423 (1982)}.

\bibitem{[Fra83]}
{S. Frauendorf and F.R. May, Phys. Lett. {\bf 125B}, 245 (1983)}.

\bibitem{[Che83a]}
{Y.S. Chen, S. Frauendorf and G.A. Leander, Phys. Rev. {\bf C28}, 2437
  (1983)}.

\bibitem{[Ham83]}
{I. Hamamoto and B. Mottelson, Phys. Lett. {\bf 127B}, 281 (1983)}.

\bibitem{[Ben84a]}
{R. Bengtsson and W. Nazarewicz, Proc. XIX Winer School on Physics, Zakopane,
  1984, ed. by Z. Stachura, Report IFJ No. 1268, p. 171}.

\bibitem{[Ben88]}
{T. Bengtsson, S. \AA berg and I. Ragnarsson, Phys. Lett. {\bf 208B}, 39
  (1988)}.

\bibitem{[Jan91]}
{R.V.F. Janssens and T.L. Khoo, Ann. Rev. Nucl. Part. Sci. {\bf 41}, 321
  (1991)}.


\bibitem{[Bak95]}
{C. Baktash, B. Haas and W. Nazarewicz, Annu. Rev. Nucl. Part. Phys. {\bf 45},
  1995, in press}.

\bibitem{[Ham87a]}
{I. Hamamoto, Phys. Lett. {\bf B 193}, 399 (1987)}.

\end{thebibliography}

\clearpage


\begin{table}[ht]
\begin{center}
\begin{tabular}{ccrrrrrrrrrr}
\multicolumn{2}{c}{Nucleus} & \multicolumn{3}{c}{Oblate} 
& \multicolumn{3}{c}{Prolate}
& $\Delta E_{\rm op}$ & Q$_{\rm o}$ & Q$_{\rm p}$ & 
Q$_{\rm exp}$ \\
Z & A & $\beta_2$ & $\beta_4$ & $\beta_6$ &
$\beta_2$ & $\beta_4$ & $\beta_6$ & MeV & eb & eb & eb \\
 \hline
 34 & 56 &  -0.22 &   0.04 &   0.08 &   0.19 &  -0.07 &   0.01 &   0.29 &  -1.64 &   1.56 &   \\
    & 58 &  -0.25 &   0.05 &   0.09 &   0.21 &  -0.07 &   0.01 &  -0.30 &  -1.91 &   1.76 &   \\
    & 60 &  -0.28 &   0.06 &   0.09 &   0.22 &  -0.08 &   0.02 &  -1.14 &  -2.15 &   1.86 &   \\
    &    &        &        &        &   0.31 &   0.08 &   0.04 &   0.10 &        &   3.25 &  \\
    & 62 &  -0.29 &   0.05 &   0.10 &   0.22 &  -0.10 &   0.03 &  -1.36 &  -2.22 &   1.88 &   \\
    &    &        &        &        &   0.32 &   0.06 &   0.03 &  -0.20 &        &   3.31 &  \\
    & 64 &  -0.29 &   0.03 &   0.11 &   0.26 &  -0.07 &   0.03 &  -0.97 &  -2.20 &   2.28 &   \\
    &    &        &        &        &   0.33 &   0.03 &   0.02 &   0.08 &        &   3.27 &  \\
    & 66 &  -0.28 &   0.01 &   0.09 &   0.27 &  -0.07 &   0.03 &  -0.57 &  -2.14 &   2.47 &   \\
    & 68 &  -0.28 &  -0.02 &   0.08 &   0.29 &  -0.08 &   0.02 &  -0.43 &  -2.13 &   2.61 &   \\
   & & & & & & & & & & & \\[-3mm]
 36 & 56 &  -0.24 &   0.05 &   0.07 &   0.16 &  -0.05 &   0.00 &  -0.07 &  -1.93 &   1.40 &   \\
    & 58 &  -0.30 &   0.08 &   0.11 &   0.30 &   0.04 &   0.05 &  -0.55 &  -2.44 &   3.10 &   \\
    & 60 &  -0.31 &   0.07 &   0.10 &   0.32 &   0.04 &   0.04 &  -0.86 &  -2.52 &   3.41 &   \\
    & 62 &  -0.32 &   0.06 &   0.11 &   0.33 &   0.03 &   0.03 &  -0.81 &  -2.63 &   3.51 &   \\
    & 64 &  -0.32 &   0.05 &   0.10 &   0.33 &   0.01 &   0.03 &  -0.39 &  -2.61 &   3.52 &   \\
    & 66 &  -0.31 &   0.03 &   0.10 &   0.33 &  -0.02 &   0.04 &  -0.06 &  -2.56 &   3.39 &   \\
    & 68 &  -0.31 &   0.00 &   0.09 &   0.32 &  -0.05 &   0.04 &   0.11 &  -2.52 &   3.28 &   \\
    & 70 &  -0.31 &  -0.01 &   0.08 &   0.31 &  -0.07 &   0.03 &   0.06 &  -2.59 &   3.16 &   \\
    & 72 &  -0.32 &   0.00 &   0.06 &   0.30 &  -0.06 &   0.02 &   0.25 &  -2.71 &   3.01 &   \\
   & & & & & & & & & & & \\[-3mm]
 38 & 56 &  -0.11 &  -0.03 &   0.00 &   0.27 &   0.04 &   0.04 &  -0.85 &  -0.92 &   2.96 &   \\
    & 58 &  -0.28 &   0.05 &   0.07 &   0.33 &   0.04 &   0.04 &   0.41 &  -2.40 &   3.70 &   \\
  & 60 &  -0.33 &  0.07 &  0.10 &  0.34 &  0.04 &  0.03 &  0.49 &  -2.81 &   3.84 & 3.12\\
  & 62 &  -0.34 &   0.07 &   0.11 &   0.35 &   0.03 &   0.02 &   0.64 & -2.96 & 3.98 &
3.32  \\
    & 64 &  -0.34 &   0.06 &   0.11 &   0.36 &   0.01 &   0.01 &   1.10 &  -2.98 &   4.04 &   \\
    & 66 &  -0.32 &   0.03 &   0.09 &   0.35 &  -0.01 &   0.02 &   1.48 &  -2.84 &   3.97 &   \\
    & 68 &  -0.30 &  -0.01 &   0.07 &   0.35 &  -0.04 &   0.03 &   1.48 &  -2.64 &   3.87 &   \\
    & 70 &  -0.30 &  -0.03 &   0.06 &   0.35 &  -0.05 &   0.02 &   1.38 &  -2.65 &   3.88 &   \\
 & 72 &  -0.32 &  -0.01 &   0.06 &   0.36 &  -0.05 &   0.02 &   1.22 &  -2.83 &   4.06 & \\
   & & & & & & & & & & & \\[-3mm]
\end{tabular}
\end{center}
\caption{\label{table_prolate-oblate}
Calculated 
equilibrium deformations  $\beta_2, \beta_4$, and $\beta_6$ of 
the  ground-state and excited minima in
neutron-rich  
even-even  Se-Cd isotopes. 
The   energy differences between the local minima, 
and the corresponding quadrupole moments are also shown, together
with experimental quadrupole moments~\protect\cite{[Ram87]}.
}
\end{table}
\setcounter{table}{0}
\newpage

\begin{table}[ht]
\begin{center}
\begin{tabular}{ccrrrrrrrrrr}
\multicolumn{2}{c}{Nucleus} & \multicolumn{3}{c}{Oblate} 
& \multicolumn{3}{c}{Prolate}
& $\Delta E_{\rm op}$ & Q$_{\rm o}$ & Q$_{\rm p}$ & 
Q$_{\rm exp}$ \\
Z & A & $\beta_2$ & $\beta_4$ & $\beta_6$ &
$\beta_2$ & $\beta_4$ & $\beta_6$ & MeV & eb & eb & eb \\
 \hline
 40 & 56 &        &        &        &   0.00 &   0.00 &   0.00 &  &     &   0.00 & 0.74 \\
  & 58 &  -0.17 &  -0.02 &  -0.01 &   0.29 &   0.05 &   0.02 & -0.05 & -1.60 & 3.40 &
\\
 & 60 &  -0.21 &  -0.01 &   0.01 &   0.34 & 0.03 &  0.00 &  0.85 & -2.01 & 4.09 &3.01 \\
 & 62 &  -0.24 &   0.00 &   0.01 &   0.36 &  0.02 & -0.01 & 1.51 & -2.25 & 4.37 &4.01 \\
    & 64 &  -0.25 &  -0.02 &   0.02 &   0.37 &   0.00 &  -0.01 &   1.67 &  -2.33 &   4.45 &   \\
    & 66 &  -0.24 &  -0.04 &   0.01 &   0.37 &  -0.02 &  -0.01 &   1.42 &  -2.25 &   4.37 &   \\
    & 68 &  -0.25 &  -0.06 &   0.01 &   0.36 &  -0.05 &   0.00 &   1.02 &  -2.26 &   4.23 &   \\
    & 70 &  -0.25 &  -0.07 &   0.01 &   0.36 &  -0.06 &   0.00 &   0.84 &  -2.28 &   4.24 &   \\
    & 72 &  -0.22 &  -0.08 &  -0.01 &   0.38 &  -0.06 &  -0.01 &   0.45 &  -2.05 &   4.59 &   \\
    & 74 &  -0.18 &  -0.09 &  -0.03 &   0.37 &  -0.05 &  -0.01 &  -0.90 &  -1.70 &   4.50 &   \\
   & & & & & & & & & & & \\[-3mm]
 42& 58 &  -0.18 &  -0.02 & 0.00 & 0.00 &   0.00 & 0.00 & 0.01 &  -1.79 &0.00 & 2.28  \\
   &    &        &        &      & 0.20 &   0.05 & 0.01 & 0.05 &    &   2.42 &  \\
 & 60 &  -0.21 &  -0.02 &   0.00 &   0.27 &   0.05 &   0.01 &   0.22 & -2.08 & 3.46 &3.26\\
 &    &        &        &        &   0.34 &   0.02 &  -0.03 &   0.30 &      &   4.27 &  \\
 & 62 &  -0.23 &  -0.03 &  0.00 &  0.35 &  0.03 & -0.02 &  0.60 & -2.23 & 4.51 & 3.29 \\
 & 64 &  -0.24 &  -0.05 &  0.00 &  0.38 &  0.02 & -0.03 &  0.58 & -2.29 & 4.91 & 3.62 \\
 & 66 &  -0.24 &  -0.06 &  0.00 &  0.38 & -0.01 & -0.02 &  0.13 & -2.32 & 4.86 & 3.68 \\
 & 68 &  -0.25 &  -0.08 &   0.00 &   0.33 &  -0.04 &   0.00 &  -0.49 &  -2.40 &   4.15 &   \\
    & 70 &  -0.25 &  -0.09 &   0.00 &   0.33 &  -0.06 &   0.00 &  -0.91 &  -2.36 &   4.02 &   \\
    & 72 &  -0.23 &  -0.09 &  -0.01 &   0.28 &  -0.05 &   0.00 &  -1.42 &  -2.27 &   3.49 &   \\
    &    &        &        &        &   0.42 &  -0.05 &  -0.01 &  -0.07 &        &   5.45 &  \\
    & 74 &  -0.21 &  -0.10 &  -0.03 &   0.21 &  -0.03 &   0.01 &  -2.22 &  -2.00 &   2.67 &   \\
   & & & & & & & & & & & \\[-3mm]
 44 & 60 &  -0.21 &  -0.02 &  -0.01 &   0.22 &   0.05 &  0.00 &  0.62 & -2.18 & 2.88 &
2.91 \\
 & 62 &  -0.23 &  -0.03 &  -0.01 &   0.25 &   0.04 &  -0.01 &   0.45 &  -2.40 & 3.40 &   \\
 & 64 &  -0.24 &  -0.05 &   0.00 &   0.26 &   0.03 &  -0.01 &   0.15 &  -2.45 &   3.54 &
3.22   \\
 & 66 &  -0.24 &  -0.07 &   0.00 &   0.26 &   0.01 &  -0.01 &  -0.36 &  -2.47 & 3.41 & 3.34  \\
 & 68 &  -0.25 &  -0.09 &   0.00 &   0.25 &  -0.02 &   0.00 &  -0.88 &  -2.51 & 3.30 & 3.36  \\
 & 70 &  -0.25 &  -0.09 &   0.00 &   0.24 &  -0.03 &   0.00 &  -1.19 &  -2.52 &   3.17 &   \\
 & 72 &  -0.24 &  -0.10 &  -0.01 &   0.22 &  -0.04 &   0.00 &  -1.43 &  -2.41 &   2.89 &   \\
 & 74 &  -0.21 &  -0.11 &  -0.03 &   0.17 &  -0.02 &   0.00 &  -1.58 &  -2.12 &   2.21 &   \\
 & 76 &  -0.18 &  -0.11 &  -0.04 &   0.08 &   0.00 &   0.00 &  -1.31 &  -1.90 &   1.11 &   \\
 &    &        &        &        &   0.51 &   0.04 &   0.05 &   5.08 &  &  8.05 & \\
\end{tabular}
\end{center}
\caption{(Continued)}
\end{table}

\setcounter{table}{0}
\newpage

\begin{table}[ht]
\begin{center}
\begin{tabular}{ccrrrrrrrrrr}
\multicolumn{2}{c}{Nucleus} & \multicolumn{3}{c}{Oblate} 
& \multicolumn{3}{c}{Prolate}
& $\Delta E_{\rm op}$ & Q$_{\rm o}$ & Q$_{\rm p}$ & 
Q$_{\rm exp}$ \\
Z & A & $\beta_2$ & $\beta_4$ & $\beta_6$ &
$\beta_2$ & $\beta_4$ & $\beta_6$ & MeV & eb & eb & eb \\
 \hline
 46 & 56 &        &        &        &   0.02 &   0.01 &   0.00 &   &  &     0.21 & 2.15 \\
    & 58 &        &        &        &   0.14 &   0.03 &  -0.01 &   &  &     1.89 & 2.32 \\
    & 60 &        &        &        &   0.18 &   0.03 &  -0.01 &   &   &    2.42 & 2.57 \\
    & 62 &  -0.23 &  -0.01 &  -0.01 &   0.21 &   0.04 &  -0.01 &   1.00 &  -2.60 & 2.97 &2.76 \\
    & 64 &  -0.24 &  -0.03 &  -0.01 &   0.22 &   0.03 &  -0.01 &   0.74 &  -2.63 & 3.11 &2.96 \\
    & 66 &  -0.24 &  -0.05 &   0.00 &   0.21 &   0.00 &   0.00 &   0.37 &  -2.64 & 2.91 &2.52 \\
    & 68 &  -0.24 &  -0.07 &   0.00 &   0.21 &  -0.02 &   0.00 &   0.00 &  -2.66 & 2.82 &1.84 \\
    & 70 &  -0.24 &  -0.08 &   0.00 &   0.20 &  -0.03 &   0.00 &  -0.20 &  -2.65 & 2.77 &2.40 \\
    & 72 &  -0.23 &  -0.08 &  -0.01 &   0.18 &  -0.03 &   0.00 &  -0.24 &  -2.54 &   2.47 &   \\
    & 74 &  -0.19 &  -0.09 &  -0.03 &   0.14 &  -0.02 &   0.00 &  -0.16 &  -2.12 &   1.91 &   \\
    & 76 &  -0.13 &  -0.07 &  -0.03 &   0.12 &  -0.01 &   0.00 &  -0.07 &  -1.56 &   1.61 &   \\
    &    &        &        &        &   0.51 &   0.03 &   0.06 &   7.52 &        &   8.40 &  \\
    & 78 &   0.00 &   0.00 &   0.00 &   0.54 &   0.03 &   0.05 &  -9.25 &   0.03 &   9.07 &   \\
   & & & & & & & & & & & \\[-3mm]
 48 & 56 &        &        &        &   0.00 &   0.00 &   0.00 &   &  &     0.00 &  \\
    & 58 &        &        &        &   0.03 &   0.00 &   0.00 &   &  &     0.43 & 2.03 \\
    & 60 &  -0.04 &   0.00 &   0.00 &   0.12 &   0.01 &   0.00 &   0.24 &  -0.50 & 1.66 &2.08 \\
    & 62 &  -0.07 &   0.01 &   0.00 &   0.15 &   0.00 &   0.00 &   0.42 &  -0.90 & 2.07 &2.13 \\
    & 64 &  -0.09 &   0.01 &   0.00 &   0.17 &   0.01 &   0.00 &   0.52 &  -1.18 & 2.49 &2.26 \\
    & 66 &  -0.10 &   0.00 &   0.00 &   0.18 &  -0.01 &   0.01 &   0.57 &  -1.34 & 2.61 &2.35 \\
    & 68 &  -0.11 &  -0.01 &   0.00 &   0.17 &  -0.03 &   0.01 &   0.56 &  -1.37 & 2.41 &2.37 \\
    & 70 &  -0.10 &  -0.01 &   0.00 &   0.16 &  -0.03 &   0.01 &   0.48 &  -1.25 &   2.31 &   \\
    & 72 &  -0.08 &  -0.02 &   0.00 &   0.14 &  -0.03 &   0.01 &   0.24 &  -1.09 &   1.96 &   \\
    & 74 &  -0.07 &  -0.02 &  -0.01 &   0.09 &  -0.02 &   0.00 &   0.07 &  -0.90 &   1.28 &   \\
    & 76 &        &        &        &   0.00 &   0.00 &   0.00 &   &    &  -0.03 &  \\[1mm]
\end{tabular}
\end{center}
\caption{(Continued)}
\end{table}

\begin{table}
\begin{center}
\begin{tabular}{cccccccccc}%
  &   & \multicolumn{4}{c}{PNP} & \multicolumn{4}{c}{BCS}\\
Z & N & $\beta_2$ & $\beta_4$ & $\gamma$ & $Q_{\rm cal}$ 
&$\beta_2$&$\beta_4$&$\gamma$&$Q_{\rm cal}$\\
\hline
 42& 58&   0.210&   0.011& -32.726&   3.008&  0.162&   0.001& -51.801&   2.103\\
   & 60&   0.300&   0.035&   0.000&   3.973&  0.243&   0.019& -20.538&   3.547\\
   & 62&   0.316&   0.017& -19.046&   4.737&  0.291&   0.021& -14.055&   4.261\\
   & 64&   0.335&   0.016& -19.695&   5.116&  0.308&   0.015& -16.866&   4.619\\
   & 66&   0.328&   0.001& -21.119&   5.008&  0.306&   0.005& -18.690&   4.629\\
 44& 58&   0.231&   0.015& -25.030&   3.541&  0.125&   0.013&   0.000&   1.606\\
   & 60&   0.252&   0.019& -23.157&   3.940&  0.184&   0.020&   0.000&   2.460\\
   & 62&   0.275&   0.003& -19.507&   4.271&  0.245&   0.003& -18.232&   3.756\\
   & 64&   0.284&   0.014& -19.875&   4.533&  0.263&   0.013& -21.254&   4.185\\
   & 66&   0.283&   0.004& -21.870&   4.550&  0.272&   0.012& -23.660&   4.413\\
\end{tabular}
\end{center}
\caption{ \label{table_threeDminima}
Equilibrium deformations
$\beta_2$, $\gamma$, and $\beta_4$ and corresponding quadrupole 
moments (in eb) for the Mo and Ru isotopes calculated
in the PNP (left) and BCS (right) variants.
}
\end{table}
\newpage

\begin{figure}
\caption{Single particle 
neutron (top) and proton (bottom)
levels of the deformed Woods-Saxon potential
as a function of quadrupole deformation $\beta_2$.
Positive (negative) parity states are indicated by
solid (dotted) lines, and the spherical and deformed shell and
subshell closures are indicated.
Calculations are performed with the
Woods-Saxon set of parameters of
Ref.~\protect\cite{[Naz88]} 
corresponding to  the  central nucleus $^{100}$Zr.
\label{WSLevel}
}
\end{figure}

\begin{figure}
\caption{Potential-energy curves 
for the even-even neutron-rich isotopes of Kr, Sr, and Zr 
obtained in the FRLD/PNP  (left) and FRLD/BCS 
(right) variant of calculations.
At each value of $\beta_2$,
the energy has been minimized with respect to deformations $\beta_4$
and $\beta_6$.
\label{PECKrZr}
}
\end{figure}

\begin{figure}
\caption{Same as in Fig.~\protect\ref{PECKrZr} except for
the even-even neutron-rich isotopes of Mo, Ru, and Pd.
\label{PECMoPd}
}
\end{figure}

\begin{figure}
\caption{Total energy surfaces in the ($\beta_2, \gamma$)-plane
for $^{100,102}$Zr, in  the
LD/PNP (left) and 
 LD/BCS (right) 
 variant of calculations.
At each mesh point ($\beta_2$, $\gamma$) 
the total energy has been minimized with respect to $\beta_4$. 
The distance between thick contour lines is 1\,MeV, while
between the thin contour line it is
 250\,keV.
\label{PESZra}}
\end{figure}

\begin{figure}
\caption{Same as in Fig. ~\protect\ref{PESZra} except
 for 
$^{104,106}$Zr.
\label{PESZrb}}
\end{figure}

\begin{figure}
\caption{Same as in Fig. ~\protect\ref{PESZra} except
 for 
$^{100,102}$Mo.
\label{PESMoa}}
\end{figure}

\begin{figure}
\caption{Same as in Fig. ~\protect\ref{PESZra} except
 for 
$^{104,106}$Mo.
\label{PESMob}}
\end{figure}

\begin{figure}
\caption{Same as in Fig. ~\protect\ref{PESZra} except
for 
$^{104,106}$Ru.
\label{PESRua}}
\end{figure}

\begin{figure}
\caption{Same as in Fig. ~\protect\ref{PESZra} except
 for 
$^{108,110}$Ru.
\label{PESRub}}
\end{figure}

\begin{figure}
\caption{Total routhian surfaces in the ($\beta_2, \gamma$)-plane
for the ($\pi$=+, $r$=1) quasi-particle vacuum configuration  of 
$^{100}$Sr at four values of rotational
frequency: $\hbar\omega$=0.3, 0.6, 0.9, and 1.2\,MeV.
At each ($\beta_2$, $\gamma$)  point
the total routhian has been minimized with respect to
hexadecapole deformation
 $\beta_4$. 
The distance between thick contour lines is 1\,MeV, while
between the thin contour line it is
 250\,keV.
The angular momentum values at local minima are indicated.
\label{TRS100Sr}}
\end{figure}

\begin{figure}
\caption{Same as in Fig.~\protect\ref{TRS100Sr} except for
$^{102}$Zr.
\label{TRS102Zr}}
\end{figure}

\begin{figure}
\caption{Same as in Fig.~\protect\ref{TRS100Sr} except for
$^{108}$Ru.
\label{TRS108Ru}}
\end{figure}

\begin{figure}
\caption{Summary of calculated 
equilibrium deformations of  yrast and near-yrast
($\pi$=+, $r$=1) bands in
$^{96,98,100,102,104}$Sr.
The rotational frequency is varied
from $\hbar\omega$=0 to $\hbar\omega$=1.5\,MeV, with steps
 of $\Delta\omega$=0.07 MeV$/\hbar$. If the two minima 
corresponding to the TRS's calculated
at $\omega$ and $\omega$+$\Delta\omega$
have the same intrinsic configuration, and the change in 
equilibrium deformation is small, 
$\Delta (\beta_2\cos{\gamma}) \le 0.1$, 
they are assumed to belong to the same rotational band 
and they are connected by a solid line.  Whenever  feasible,
the direction of  increasing $\omega$ is marked by an arrow.
The symbols indicating the intrinsic configuration are
put at the lowest and highest frequency at which 
the band is calculated, and the corresponding spin range is shown by numbers.
The legend displays band labels. 
Superdeformed bands and high-spin bands with $\gamma$$>$0
 are classified according to the number of occupied
high-${\cal N}$ intruder levels 
($\cal N$ = 6 and 5 for neutrons and protons, respectively).
Other band structures are classified according
to the number of aligned high-$j$ quasi-particles:
 {\nh}, {\pg}, and {\na}.
\label{BCSTRA-Sr}
}
\end{figure}

\begin{figure}
\caption{Same as in Fig.~\protect\ref{BCSTRA-Sr} except for
$^{98,100,102,104,106}$Zr.
\label{BCSTRA-Zr}}
\end{figure}

\begin{figure}
\caption{Same as in Fig.~\protect\ref{BCSTRA-Sr} except for
$^{100,102,104,106,108}$Mo.
\label{BCSTRA-Mo}}
\end{figure}

\begin{figure}
\caption{Same as in Fig.~\protect\ref{BCSTRA-Sr} except for
$^{102,104,106,108,110}$Ru.
\label{BCSTRA-Ru}}
\end{figure}

\begin{figure}
\caption{
Single quasi-particle routhian diagram for 
$N$=62 (top) and $Z$=40 (bottom) at
$\beta_2$=0.3 and  $\gamma$=0$^\circ$. This diagram is typical for
the  well-deformed prolate bands in the heavy-Zr region.
The parity and signature of individual levels are indicated 
in the following way:
$\pi$=+, $r$=$-i$ - solid line;
$\pi$=+, $r$=$+i$ - dotted line;
$\pi$=--, $r$=$-i$ -  dot-dashed line;
$\pi$=--, $r$=$+i$ - dashed line.
The thin line indicates the Fermi energy.
\label{Rou102Zr}
}\end{figure}

\begin{figure}
\caption{Experimental  and calculated
 angular momentum  alignment  for
yrast bands of well-deformed $^{100,102}$Zr. 
Experimental data are taken from Ref. \protect\cite{[Dur96]}.
The diabatic bands have not been extracted in the calculation.
Therefore, the band crossing region is not reproduced theoretically.
\label{IWNormal}
}\end{figure}

\begin{figure}
\caption{
Single quasi-particle routhian diagram for 
$N$=64 (top) and $Z$=44 (bottom) at
$\beta_2$=0.23 and  $\gamma$=--30$^\circ$. This diagram is typical for
the  triaxial bands in the heavy-Mo region.
The line characteristics have the same interpretation as in
Fig.~\protect\ref{Rou102Zr}.
\label{Rou108Ru}
}\end{figure}

\begin{figure}
\caption{Experimental  and calculated
angular momentum alignment
for triaxial yrast line in   $^{108}$Ru.
Experimental data are taken from Ref.~\protect\cite{[Lu95]}.
The diabatic bands have not been extracted in the calculation.
Therefore, the band crossing region is not reproduced theoretically.
\label{IWTriaxial}
}\end{figure}

\begin{figure}
\caption{
Single-particle routhian diagram for neutrons (top)
and protons (bottom) at  $\beta_2$=0.45 and  $\gamma$=0$^\circ$, 
 characteristic of SD configurations
in the heavy Zr region.
The line convention is the same as in Fig.~\protect\ref{Rou102Zr}.
\label{RouSD}
}
\end{figure}

\begin{figure}
\caption{
Calculated kinematic moments of inertia versus rotational frequency 
for superdeformed bands in even-even nuclei around $^{104}$Mo. 
The symbols indicate the same configurations as in
 Figs.~\protect\ref{BCSTRA-Sr}-\protect\ref{BCSTRA-Ru}.
\label{IWSuper}
}\end{figure}

\newpage
\begin{center}
{{\hyphenation{retains}
$$\vbox{\hsize=6.2in\parindent=0pt\hbadness=10000
        \baselineskip=7pt\footnotesize
``The submitted manuscript has been authored by a
contractor of the U.S.\ Government under contract
DE-AC05-96OR22464.  Accordingly, the U.S.\ Government
retains a nonexclusive, royalty-free license to
publish or reproduce the published form of this
contribution, or allow
others to do so, for U.S.\ Government purposes.''\par}
$$}}
\end{center}

\end{document}